\documentclass[twocolumn,preprintnumbers,amsmath,amssymb,aps,prl]{revtex4-2}

\usepackage[pdftex]{graphicx}
\usepackage{dcolumn}
\usepackage{bm}
\usepackage{float}
\usepackage{hyperref}
\usepackage{epstopdf}
\usepackage{mathrsfs}
\usepackage{amsmath}
\usepackage{graphicx}
\usepackage{multirow}
\usepackage{xr-hyper}

\newcommand{\ket}[1]{\left| #1 \right>}
\newcommand{\bra}[1]{\left< #1 \right|}

\newcommand{\Eq}[1]{Eq.\,\eqref{#1}}

\hypersetup{
            colorlinks=true,
            linkcolor=blue,
            anchorcolor=blue,
            citecolor=blue}
\begin{document}

\title{Field-induced Berry connection and planar  Hall effect  in tilted Weyl semimetals}
\author{YuanDong Wang$^{1,2}$}

\author{Zhen-Gang Zhu$^{1,2,3}$}
\email{zgzhu@ucas.ac.cn}

\author{Gang Su$^{2,3,4}$}
\email{gsu@ucas.ac.cn}

\affiliation{$^{1}$ School of Electronic, Electrical and Communication Engineering, University of Chinese Academy of Sciences, Beijing 100049, China.\\
$^{2}$ School of Physical Sciences, University of Chinese Academy of Sciences, Beijing 100049, China. \\
$^{3}$ CAS Center for Excellence in Topological Quantum Computation, University of Chinese Academy of Sciences, Beijing 100049, China.\\
$^{4}$ Kavli Institute for Theoretical Sciences, University of Chinese Academy of Sciences, Beijing 100190, China.
 }
\begin{abstract}
We propose the linear and nonlinear planar Hall effect (PHE) in tilted Weyl semimetals in the presence of an in-plane magnetic and electric field, where the field-induced Berry connection (FBC) plays a key role.
We show that the PHE is ascribed to the quantum metric, distinct from the well-known chiral anomaly-induced PHE arising from the Berry curvature.  Using a tilting vector to describe the model, we demonstrate the constrains on the linear and nonlinear PHE by the tilting directions.
The linear PHE is intrinsic that is determined by the topological properties of energy bands, whereas the nonlinear PHE is extrinsic.
The predicted linear and nonlinear PHE are inherently different from others and may shed light on a deeper understanding on transport nature of the tilted Weyl semimetals.
\end{abstract}
\maketitle

\textit{{\color{blue}Introduction.-}}
Three-dimensional topological Weyl semimetals (WSMs), a new state of quantum matter with gapless spectrum at bulk nodal points and spin non-degenerate open Fermi arc states on the surface, has attracted much recent interest \cite{PhysRevB.83.205101, PhysRevX.5.011029, PhysRevX.5.031013, xu2015, xu2015-2, doi:10.1146/annurev-conmatphys-031016-025225, doi:10.1146/annurev-conmatphys-031016-025458, RevModPhys.90.015001}. The Weyl monopoles hosted by topological semimetals, as sources or sinks of the Berry curvature may lead to a number of nontrivial transport effects, including the anomalous Hall effect \cite{PhysRevB.83.205101, PhysRevB.84.075129, PhysRevLett.107.127205, PhysRevLett.107.186806, PhysRevLett.126.106601}, in which a transverse charge current proportional to the Berry curvature  is generated in response to a longitudinal electric field without external magnetic fields,  the ``chiral anomaly" \cite{PhysRevB.88.104412, PhysRevLett.113.247203, PhysRevB.88.115119, PhysRevX.4.031035} that breaks the chiral symmetry leading to the nonconservation of chiral charges, etc. Until now, the negative longitudinal magnetoresistance and the planar Hall effect (PHE) are the most remarkable phenomena induced by the chiral anomaly 
\cite{Burkov_2015,PhysRevB.95.245128,PhysRevB.89.195137,PhysRevLett.119.176804,PhysRevB.96.041110,PhysRevB.102.121105}.
For the PHE, a net charge current $\bm{J}\propto (\bm{E}\cdot\bm{B})\bm{B}$ \cite{PhysRevLett.109.181602, PhysRevLett.109.162001,PhysRevLett.107.021601,PhysRevB.91.115203} can be induced due to chiral anomaly and non-conservation of density of electrons at an individual node when the magnetic field and the  electric field are nonorthogonal (i.e., $\bm{E}\cdot \bm{B}\neq 0$) \cite{PhysRevB.88.104412}.
%

The Berry curvature can be recognized as the imaginary part of a quantum geometric tensor \cite{berry1984quantal, berry1989geometric}, whereas the real part of this tensor is another important geometric quantity called quantum metric \cite{provost1980riemannian}, which allows us to measure the distance between quantum states. The concept of the quantum metric was well applied in quantum information theory \cite{PhysRevA.54.1844, PhysRevLett.78.2275, PhysRevLett.113.140401, PhysRevA.71.032313, leibfried2003experimental}, and it begins to attract attention  in condensed matter physics \cite{peotta2015superfluidity, PhysRevLett.117.045303, PhysRevLett.121.020401, PhysRevLett.122.227402, torma2021superfluidity}.
Recently, the quantum metric is found to have relationship with field-induced Berry connection (FBC) \cite{PhysRevLett.112.166601, PhysRevB.91.214405} by invoking a generalized semiclassical theory. Therefore, it is quite imperative to know whether a measurable effect can exist to reflect the character of the quantum metric. 


In this paper, we propose a new intrinsic linear and extrinsic nonlinear PHE inherent to the quantum metric linked via FBC.  This may provide a direct method unveiling the profound connection between transport in condensed matters and the quantum metric. Since the linear effect should be dominant in the transport when the linear and nonlinear effects occur simultaneously, the new proposed intrinsic linear PHE is crucial. The nonlinear PHE proposed here is extrinsic  that differs from others. Moreover, we point out the case in which the linear PHE disappears whereas the nonlinear PHE is dominant, which is also enlightening to topological nonlinearity.



\textit{{\color{blue}Field-induced Berry connection and the corresponding current.-}}
We 
start from the Berry curvature of the $n$th band which is defined by
\begin{equation}
\bm{\Omega}_{n}(\bm{k})=\bm{\nabla}_{\bm{k}}\times \bm{\mathcal{A}}_{n}(\bm{k}),
\end{equation}
where $\bm{\mathcal{A}}_{n}(\bm{k})=i\bra{u_{n\bm{k}}}\bm{\nabla}_{\bm{k}}\ket{u_{n\bm{k}}}$ is the intraband Berry connection with $\ket{u_{n\bm{k}}}$ the periodic part of the $n$th Bloch state. In the presence of electric field $\bm{E}$ and magnetic field $\bm{B}$, the Berry connection acquires a gauge-invariant correction besides $\bm{\mathcal{A}}_{n}(\bm{k})$, which is written as
\begin{equation}
\bm{\mathcal{A}}^{\prime}_{n}=\bm{\mathcal{A}}^{\prime}_{n}(E)+\bm{\mathcal{A}}^{\prime}_{n}(B).
\end{equation}
As a result, the Berry curvature is extended to $\tilde{\bm{\Omega}}_n=\bm{\Omega}_n +\bm{\Omega}_{n}^\prime$ with $\bm{\Omega}_{n}^\prime =\bm{\nabla}\times \bm{\mathcal{A}}_{n}^\prime $. The E-FBC can be written as $\bm{\mathcal{A}}^{\prime}_{n}(E)=\overleftrightarrow{\bm{G}}_{n}\bm{E}$,
where the double arrow indicates the second-rank tensor, which  is given by  \cite{PhysRevLett.112.166601, gao2019semiclassical, PhysRevLett.127.277202}
\begin{equation}\label{e-bcp}
(\overleftrightarrow{\bm{G}}_{n})^{\alpha\beta}=G^{\alpha\beta}_{n}=2\sum_{m\neq n}\frac{\text{Re}[\mathcal{A}_{nm}^{\alpha}\mathcal{A}_{mn}^{\beta}]}{\varepsilon_{n}-\varepsilon_{m}},
\end{equation}
where $\bm{\mathcal{A}}_{nm}=i\bra{u_{n\bm{k}}}\bm{\nabla}_{\bm{k}}\ket{u_{m\bm{k}}}$ is the interband Berry connection, and $\varepsilon_n$ is the unperturbed band energy where for simplicity the modification caused by the orbital magnetic moment is not considered.
The magnetic field counterpart is $\bm{\mathcal{A}}^{\prime}_{n}(B)=\overleftrightarrow{\bm{F}}_{n}\bm{B}$ \cite{PhysRevLett.112.166601, gao2019semiclassical} with
\begin{equation}\label{b-bcp}
F^{\alpha\beta}_n =\text{Im}\sum_{m,p\neq n}\frac{\mathcal{A}_{nm}^{\alpha}\epsilon^{\beta\gamma\delta}[(\varepsilon_{p}-\varepsilon_{m})\mathcal{A}^{\gamma}_{mp}+iv^{\gamma}_n \delta_{mp}]\mathcal{A}^{\delta}_{pn}}{\varepsilon_{n}-\varepsilon_{m}},
\end{equation}
in which $\bm{v}_{n}=\frac{1}{\hbar}\frac{\partial \varepsilon_{n}}{\partial \bm{k}}$ is the group velocity.
Note that the Berry  connection involves a sum over all pairs of bands with $g^{\alpha\beta}_{nm}=\text{Re}[\mathcal{A}_{nm}^{\alpha}\mathcal{A}_{mn}^{\beta}]$, which is recognised as the quantum metric for two band systems \cite{rossi2021quantum,PhysRevResearch.2.033100}.

The semiclassical equations of motion reads \cite{RevModPhys.82.1959, duval2006berry} (for brevity the sum of the band index is implied)
\begin{align}
&D(\bm{B},\tilde{\bm{\Omega}})\dot{\bm{r}}=\left[{\bm{v}}_{\bm{k}}+ \frac{e}{\hbar}\bm{E}\times \tilde{\bm{\Omega}}_{\bm{k}} + \frac{e}{\hbar}({\bm{v}}_{\bm{k}} \cdot \tilde{\bm{\Omega}}_{\bm{k}})\bm{B}\right], \label{eq-mo-r}\\
&D(\bm{B},\tilde{\bm{\Omega}})\dot{\bm{k}}=\left[-\frac{e}{\hbar}\bm{E}- \frac{e}{\hbar}{\bm{v}}_{\bm{k}}\times \bm{B} - \frac{e^2}{\hbar^2}(\bm{E} \cdot \bm{B})\tilde{\bm{\Omega}}_{\bm{k}}\right]. \label{eq-mo-k}
\end{align}
where $D(\bm{B},\tilde{\bm{\Omega}})=(1+e\bm{B}\cdot \tilde{\bm{\Omega}}/\hbar)$ is the phase volume factor \cite{PhysRevLett.95.137204, RevModPhys.82.1959} revealed in the presence of non-zero Berry curvature $\tilde{\bm{\Omega}}$ and magnetic field $\bm{B}$, which is written as $D$ in the following for a shorhand notation.  $e$ is the  (positive) elementary charge.
The anomalous velocity includes a magnetic field dependent term, which indicates that the electrons move along the magnetic field direction for one Weyl cone and along the opposite direction for the cone with the opposite chirality.

In the semiclassical framework, the total current response for a uniform system can be expressed as
\begin{equation}\label{curr}
\bm{J}=-e\int [d\bm{k}]D\dot{\bm{r}}f({\varepsilon}_{\bm{k}}),
\end{equation}
where $[d\bm{k}]$ is a shorthand notation for $d\bm{k}/(2\pi)^d$,
$f({\varepsilon}_{\bm{k}})$ is the single-particle distribution function. To calculate the current density, we use the homogeneous steady-state Boltzmann equation within the relaxation time approximation to solve the distribution function:
\begin{equation}\label{be}
\dot{\bm{k}}\cdot \bm{\nabla}_{\bm{k}}f = \frac{f_{0}-f}{\tau},
\end{equation}
where $f_0$ is the equilibrium Fermi-Dirac distribution, and $\tau$ is the  transport relaxation time. To obtain the general expressions for the nonlinear currents,
\Eq{be} is solved by expanding the distribution function up to the second order in $\bm{E}$ as $f=f_0 + f_1 + f_2$, where
\begin{align}
f_1 =& \frac{\tau}{D}\left[e\bm{E}\cdot {\bm{v}}_{\bm{k}} + \frac{e^2}{\hbar}(\bm{E} \cdot \bm{B})(\bm{\Omega}_{\bm{k}}\cdot  {\bm{v}}_{k})\right]\frac{\partial f_{0}}{\partial \varepsilon_{\bm{k}}},\label{distr-f1} \\
f_2 =& \frac{\tau}{D}\left[e\bm{E}\cdot {\bm{v}}_{\bm{k}} + \frac{e^2}{\hbar}(\bm{E} \cdot \bm{B})(\bm{\Omega}_{\bm{k}}\cdot {\bm{v}}_{k})\right]\frac{\partial f_{1}}{\partial \varepsilon_{\bm{k}}}\nonumber \\
&+\frac{\tau}{D}\left[\frac{e^2}{\hbar}(\bm{E} \cdot \bm{B})(\bm{\nabla}\times  (\overleftrightarrow{\bm{G}}\bm{E}) \cdot {\bm{v}}_{k})\right]\frac{\partial f_{0}}{\partial \varepsilon_{\bm{k}}}.\label{distr-f2}
\end{align}

\textit{{\color{blue}The intrinsic linear PHE due to FBC.-}}
Firstly, we focus on the current scales with the linear order of $\bm{B}$.
Substituting the first order distribution function \Eq{distr-f1} into \Eq{curr},
the linear current up to the order $\mathcal{O}(EB)$ in the presence of external fields is obtained as
\begin{equation}\label{j1}
\bm{J}^{(1)} = \bm{J}_{\text{AB}}^{(1)}+\bm{J}_{\text{CA}}^{(1)}+\bm{J}_{\text{FBC}}^{(1)}.
\end{equation}
The first term is attributed to the anomalous velocity due to magnetic field (the meaning of index ``AB"), which is given by
\begin{equation}
\bm{J}_{\text{AB}}^{(1)}=-\frac{e^3\tau}{\hbar}\int [d\bm{k}]\frac{\partial f_{0}}{\partial \varepsilon_{\bm{k}}}(\bm{v}_{\bm{k}} \cdot \bm{\Omega}_{\bm{k}})(\bm{E}\cdot \bm{v}_{\bm{k}})\bm{B}.
\end{equation}
This term is generally nonzero for a single cone.  When the two cones are tilted in opposite directions, the distributions in two cones shift along the opposite directions due to the modified $\bm{v}_{k}$, resulting in a finite $\bm{J}_{\text{AB}}^{(1)}$ \cite{PhysRevB.99.115121}.

The second term in \Eq{j1} stems from the effective chiral chemical potential (the third term in \Eq{eq-mo-k}) due to the chiral anomaly (for the index ``CA"), which is
\begin{equation}
\bm{J}_{\text{CA}}^{(1)}=-\frac{e^3\tau}{\hbar}\int [d\bm{k}]\frac{\partial f_{0}}{\partial \varepsilon_{\bm{k}}}(\bm{E} \cdot \bm{B})(\bm{\Omega}_{\bm{k}}\cdot \bm{v}_{k})\bm{v}_{\bm{k}}.
\end{equation}
 In the absence of the tilting, for the two-cone model of the WSMs with time-reversal symmetry, the contribution of this term is zero \cite{PhysRevLett.119.176804, PhysRevB.99.115121}. 
$\bm{J}_{\text{AB}}^{(1)}$ and $\bm{J}_{\text{CA}}^{(1)}$ are all responsible for the planar Hall effect, where the Hall current, the electric and magnetic fields are all coplanar.

With considering the FBC effect, we find a new kind of PHE, which is  given as
\begin{equation}\label{jbcp1}
\begin{aligned}
\bm{J}^{(1)}_{\text{FBC }}=-\frac{e^2}{\hbar}\int [d\bm{k}] f_{0}&\left[\bm{v}_{\bm{k}}\cdot (\bm{\nabla}\times (\overleftrightarrow{\bm{G}}\bm{E})) \bm{B} \right. \\
&\left. +\bm{E}\times (\bm{\nabla}\times (\overleftrightarrow{\bm{F}}\bm{B}))\right].
\end{aligned}
\end{equation}
Here the first term is the E-FBC component and the second term is the B-FBC component. The corresponding linear planar Hall conductivity (PHC) is given by the sum $\sigma^{\alpha\beta}_{\text{FBC}}=\sigma^{\alpha\beta}_{\text{E-FBC}}+\sigma^{\alpha\beta}_{\text{B-FBC}}$.

We suppose that the electric field is directed along the $x$ axis, $\bm{E}=E\hat{x}$, and the magnetic field lies in the $x$-$y$ plane with an angle from the $x$ axis is $\theta$, i.e., $\bm{B}=B\cos{\theta}\hat{x}+B\sin{\theta}\hat{y}$. We first investigate the E-FBC component, where the longitudinal and transverse elements are found as
\begin{eqnarray}
\sigma^{xx}_{\text{E-FBC}} &=&-\frac{e^2}{\hbar}\sum_{n}\int [d\bm{k}]f^{n}_{0}v_{n}^{\gamma}\epsilon^{\gamma\delta\zeta}\partial^{\zeta}G^{\delta x}_{n}B\cos{\theta},
\label{sig-long}\\
\sigma^{yx}_{\text{E-FBC}} &=&-\frac{e^2}{\hbar}\sum_{n}\int [d\bm{k}]f^{n}_{0}v_{n}^{\gamma}\epsilon^{\gamma\delta\zeta}\partial^{\zeta}G^{\delta x}_{n}B\sin{\theta},
\label{sig-hall}
\end{eqnarray}
where the sum over the repeated Greek indices is implicit. When $f_{0}$ attains a superscript of $n$, it means the equilibrium Fermi-Dirac distribution for the n-th band.   It is found that the longitudinal conductivity and the Hall conductivity are only differed by the angle $\theta$, for this reason we only calculate the Hall one. The B-FBC component  can be obtained by
\begin{eqnarray}
\sigma^{yx}_{\text{B-FBC}} &=& \frac{e^2}{\hbar}\sum_{n}\int [d\bm{k}]f_{0}^{n}B \left[\partial^{y}(F^{xy}_{n}\sin{\theta} + F^{xx}_{n}\cos{\theta})\right. \notag\\
&-& \left.  \partial^{x}(F^{yx}_{n}\cos{\theta} + F^{yy}_{n}\sin{\theta})\right].
\label{sig-hall-b}
\end{eqnarray}
 For E-FBC case, PHE can be vanishing when the external magnetic field is parallel to the electric field, and basically it does not disappear for any direction of the magnetic field for the B-FBC case. It is evident that  $\sigma^{xx}_{\text{B-FBC}}$  is zero.
A few general remarks on the FBC-induced linear PHE are in order. Firstly, $\sigma^{\alpha\beta}_{\text{FBC}}$ vanishes if the system is time-reversal symmetric. This can be seen from the  expression of $G_{n}^{\alpha\beta}$ and $F_{n}^{\alpha\beta}$.
The $\mathcal{T}$-symmetry leads to $\bm{\mathcal{A}}_{nm}(\bm{k})=\bm{\mathcal{A}}_{mn}(-\bm{k})$, and one obtains $G_{n}^{\alpha\beta}(\bm{k})=G_{n}^{\alpha\beta}(-\bm{k})$ and $F_{n}^{\alpha\beta}(\bm{k})=-F_{n}^{\alpha\beta}(-\bm{k})$, which gives us $\mathcal{T}$-odd $\sigma^{\alpha\beta}_{\text{FBC}}$.
Secondly,  $\bm{J}^{(1)}_{\text{FBC}}$ is independent of the relaxation time $\tau$, manifesting it as an intrinsic effect determined solely by the band structure of the materials. Thirdly, for $\sigma^{xx}_{\text{B-FBC}}$ the longitudinal component is forbidden, leaving it as a pure Hall current. While for $\sigma^{\alpha\beta}_{\text{E-FBC}}$ both of the longitudinal and Hall components are allowed.

\textit{{\color{blue}FBC-induced linear PHE in WSMs.-}}
With the help of the general expressions of the PHE arising from FBC in  \Eq{sig-hall} and \Eq{sig-hall-b}, we are now in the position to investigate the PHE with FBC in WSMs.
The low-energy effective Hamiltonian describing the WSMs can be written as \cite{RevModPhys.90.015001}
\begin{equation}\label{hami}
H =  v_{F}(s\bm{k}\cdot \bm{\sigma}+\bm{R}_{s}\cdot \bm{k}\sigma_0),
\end{equation}
where $s=\pm 1$ specify the chiralities of the Weyl nodes, $\sigma_0$ is the $2\times 2$ identity matrix, $\bm{\sigma}=(\sigma^x, \sigma^y, \sigma^z)$ are the Pauli matrices, $v_{F}$ is the Fermi velocity and $\bm{R}_{s}=(R_{s}^{x},R_{s}^{y},R_{s}^{z})$ is the tilting vector of the Weyl cone.
With the Hamiltonian \Eq{hami}, the energy dispersion for a Weyl cone of chirality $s$ is derived as
\begin{equation}
\varepsilon_{s,\pm} = v_{F}(\bm{R}_s\cdot \bm{k} \pm k),
\end{equation}
where the $+$ ($-$) sign corresponds to the conduction (valence) band and $k=\vert \bm{k} \vert  $. The group velocity is $\bm{v}_{s}=v_F(\frac{k^x}{k}+R_{s}^x,  \frac{k^y}{k}+R_{s}^y, \frac{k^z}{k}+R_{s}^z)$.

Before proceeding further, we introduce the quantum metric which is closely related to the FBC in WSMs. The proper distance between quantum states can be defined as $dr^2 = Q^{\alpha\beta} k^{\alpha}k^{\beta}$, where $Q^{\alpha\beta}=\bra{\partial^{\alpha}\psi}{\partial^{\beta}\psi}\rangle - \bra{\partial^{\alpha}\psi}{\psi}\rangle \bra{\psi}{\partial^{\beta}\psi}\rangle$ is the quantum geometric tensor \cite{PhysRevA.36.3479, provost1980riemannian, bengtsson2017geometry, cheng2010quantum, rossi2021quantum}. Its imaginary part, $\text{Im}[Q^{\alpha\beta}]=\varepsilon^{\alpha\beta\gamma}\Omega^{\gamma}$, is the antisymmetric tensor Berry curvature, and its real part is the Fubini-Study quantum metric $g^{\alpha\beta}$, $\text{Re}[Q^{\alpha\beta}]=g^{\alpha\beta}$, which is the symmetric tensor. Specially, for the two band Hamiltonian, the products of Berry connections can be rewritten by the quantum metric.
For example, the quantum metric tensor for the Weyl cone with chirality $s$ of the valence band is written as $g_{s-}^{\alpha\beta}=\text{Re}[\mathcal{A}_{s-+}^{\alpha}\mathcal{A}_{s+-}^{\beta}]$, and we obtain
\begin{equation}\label{bcp}
G^{\alpha\beta}_{s-}=2g^{\alpha\beta}_{s-}/(\varepsilon_{s-} - \varepsilon_{s+}).
\end{equation}
One observes that the quantum metric is directly related to the FBC.

To get analytical results, we restrict our discussion within the type-I WSMs, where we can only consider the conduction band when the chemical potential lies above the Weyl nodes,  and we drop the band indices in the following.

\begin{figure}[htb]
\centering
\includegraphics [width=3.4in]{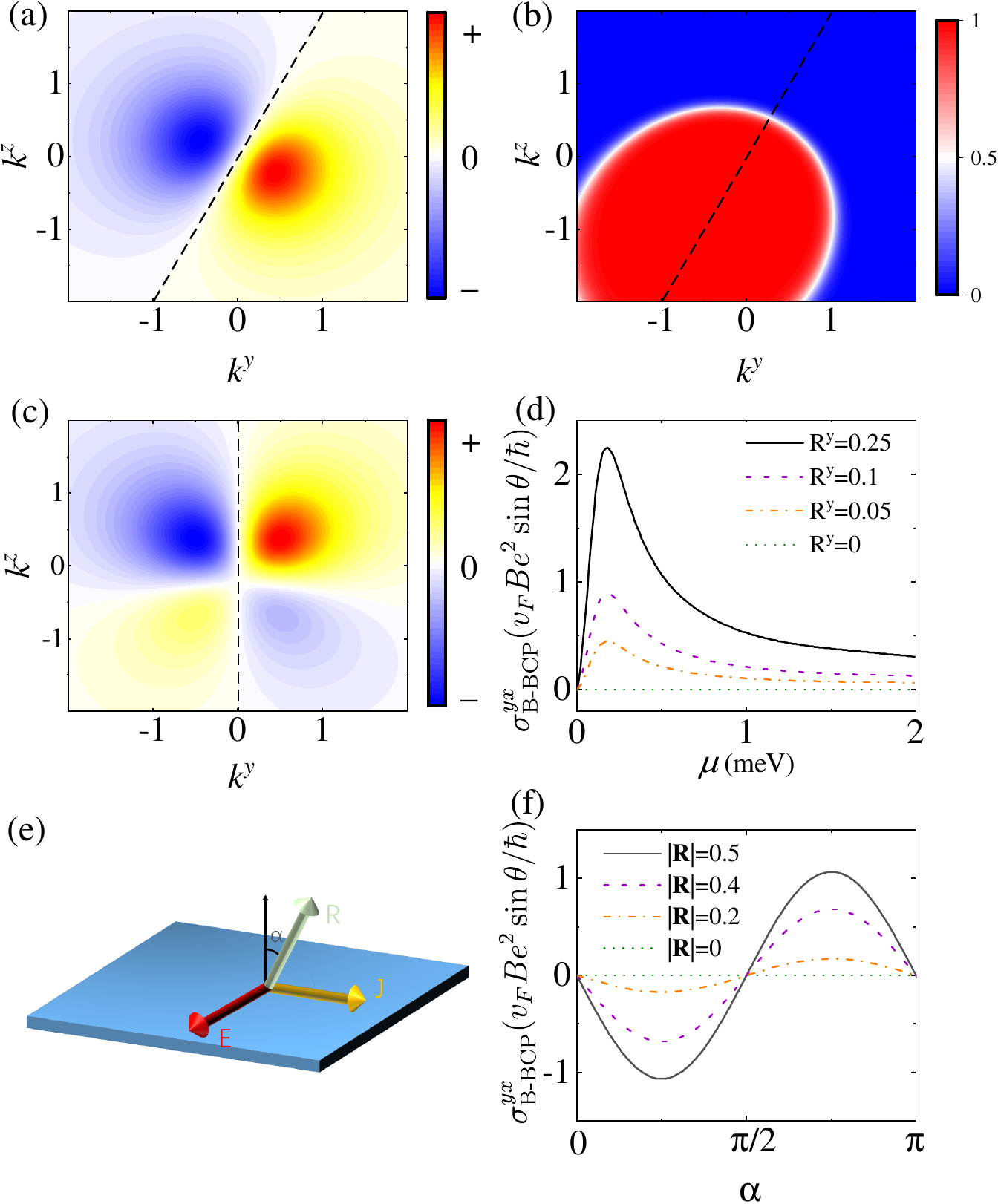}
\caption{(a) The integrand  $\Gamma^{yz}$ for $\sigma^{yx}_{\text{E-FBC}}$. (b) The equilibrium state distribution  $f_{0}$.  (c) The integrand $\Lambda^{yz}$ for $\sigma^{yx}_{\text{B-FBC}}$. For (a)-(c) the momentum plane is $k^x = 1$. (d) $\sigma^{yx}_{\text{B-FBC}}$ versus chemical potential $\mu$ for different values of $R^{y}$, $R^{x}$ is set to zero and $R^{z}$ is set to $0.5$. In the calculation, we take $k_{B}T=20$ meV. (e) Diagrammatic sketch for  the tilting vector $\bm{R}$. (f) $\sigma^{yx}_{\text{B-FBC}}$ as a function of tilting angle $\alpha$ for different values of the magnitude of $\bm{R}$.
 }\label{fig1}
\end{figure}

Let us consider the situation that a pair of Weyl cones are tilted in the same direction, in which $\bm{J}_{\text{AB}}^{(1)}$ and $\bm{J}_{\text{CA}}^{(1)}$ vanish, and $\bm{J}_{\text{FBC}}^{(1)}$ is dominated in the linear PHE. For the-same-tilting-direction case, $\bm{R}_{s}=\bm{R}$, and the PHC induced by the E-FBC is obtained as \cite{supp}
\begin{equation}
\begin{aligned}
\sigma^{yx}_{\text{E-FBC}} =-\frac{e^2}{\hbar}\int [d\bm{k}]f_{0}\Gamma^{yz}v_{F}B\sin{\theta},
\end{aligned}
\end{equation}
where
$\Gamma^{yz}=\frac{4}{k^5}(R^{z} k^y -R^{y} k^z)$.
One may observe that there is an antisymmetry mirror line with the slope  $k^{y}/k^{z}=R^{y}/R^{z}$ for $\Gamma^{yx}$ in the $k^y$-$k^z$ plane, as shown in Fig. \ref{fig1}(a), with the tilting parameters $R^{y}=0.25$ and $R^{z}=0.5$. However,  the equilibrium distribution function $f_{0}$  is symmetric with respect to the mirror line, which is depicted in Fig. \ref{fig1}(b). Combining $f_{0}$ and $\Gamma^{yz}$, it renders the cancellation of the integration in momentum space, leaving $\sigma^{yx}_{\text{E-FBC}}$ vanishes and thus we have $\sigma^{yx}_{\text{FBC}}=\sigma^{yx}_{\text{B-FBC}}$.

Now we investigate the linear PHE induced by the B-FBC. According to \Eq{jbcp1}, and making use of the two-band nature of the WSMs Hamiltonian, the complex expression of the B-FBC \Eq{b-bcp} reduces to a simple form
\begin{equation}\label{cp-f}
F^{\alpha\beta}_{s-}=\text{Re}\frac{\epsilon^{\beta\gamma\delta}v^{\gamma}_{s-}\mathcal{A}^{\alpha}_{s-+}\mathcal{A}^{\delta}_{s+-}}{\varepsilon_                                                                                                                                                                                                                                               {s-}-\varepsilon_{s+}}=\frac{\epsilon^{\beta\gamma\delta}v^{\gamma}_{s-}g^{\alpha\beta}_{s-}}{\varepsilon_                                                                                                                                                                                                                                               {s-}-\varepsilon_{s+}}.
\end{equation}
When both Weyl cones are tilted in the same direction, the linear PHE induced by the B-FBC is obtained as \cite{supp}
\begin{equation}
\begin{aligned}
\sigma^{yx}_{\text{B-FBC}} =\frac{e^2}{\hbar}\int [d\bm{k}]f_{0}(\mathcal{F}^{yxy}\sin{\theta} + \mathcal{F}^{yxx}\cos{\theta})v_{F}B,
\end{aligned}
\label{sig-hall-b-wsm}
\end{equation}
where $\mathcal{F}^{\alpha\beta\gamma}_{s}=\partial^{\alpha}F^{\beta\gamma}_{s}- \partial^{\beta}F^{\alpha\gamma}_{s}$ and we have
\begin{equation}\label{gam}
\begin{aligned}
\mathcal{F}^{yxy}=\frac{k^y (2k^z + k R^{z})}{2k^6}, \quad \mathcal{F}^{yxx}=\frac{k^x (2k^z + k R^{z})}{2k^6}.
\end{aligned}
\end{equation}
A particular case is that $R^{z}$ is zero. We thus notice that $\mathcal{F}^{yxy}$ and $\mathcal{F}^{yxx}$ are odd with respect to $k^{z}$; whereas $f_{0}$ is even, it leads to a vanishing $\sigma^{yx}_{\text{B-FBC}}$. In this case, the PHCs induced by E-FBC and B-FBC  are both zero.
Therefore a finite $R^{z}$ is required for a nonvanishing $\sigma^{yx}_{\text{B-FBC}}$, namely, a perpendicular component of the tilting vector is required. Furthermore,  $\sigma^{yx}_{\text{B-FBC}}$ vanishes if $R^{y}=R^{x}=0$.
This is because $\mathcal{F}^{yxy}$ ($\mathcal{F}^{yxx}$) is odd with respect to $k^y$ ($k^x$), as shown in Fig. \ref{fig1}(c) (where contour plot of $\mathcal{F}^{yxy}$ in $k^y$-$k^z$ plane is shown).
$R^{y}$ ($R^{x}$) breaks the reflection symmetry about the $k^y$ ($k^x$) axis for $f_{0}$, leading to a finite $\sigma^{yx}_{\text{B-FBC}}$. In Fig. \ref{fig1}(d) we plot $\sigma^{yx}_{\text{B-FBC}}$ as a function of the chemical potential $\mu$ for different values of $R^y$ ($R^{x}$ is set to zero). One observes that the magnitude of $\sigma^{yx}_{\text{B-FBC}}$ decreases for smaller $R^{y}$, as expected from symmetry analysis.

To conclude, when both Weyl cones are tilted in the same direction, the  PHE arising from FBC  is finite if the angle $\alpha$ is not zero, nor $\pi/2$, as schematically illustrated in Fig. \ref{fig1}(e), and the dependence of the tilting angle $\alpha$ for $\sigma^{yx}_{\text{B-FBC}}$ is shown in Fig. \ref{fig1}(f). It differs significantly from the linear PHE induced by the chiral anomaly where the Weyl nodes tilt along opposite directions.
%
Recently, an unconventional planar Hall effect was reported in the Weyl semimetal material $\mathrm{ZrTe_5}$ \cite{ge2020unconventional}. For which a remarkable discovery is the nonzero Hall conductivity when the in-plane magnetic field is parallel or perpendicular to the current, which is not included in previous theoretical and experimental studies (where such a conductivity disappears for the parallel case) but consistent with the planar Hall effect we proposed here.
We anticipate this new kind of linear PHE can be observed in type-I WSMs such as TaAs \cite{PhysRevX.5.011029, PhysRevLett.117.146401} with strain-controlled tilting.

\begin{table*}[hptb]
	\renewcommand\arraystretch{2}
	\caption{List of constraints on the linear and second-order PHCs  by the tilting directions of the Weyl nodes. The allowed (forbidden) PHCs  are indicated by $\checkmark$ ($\times$). }\label{tab1}
	\begin{tabular*}{17cm}{@{\extracolsep{\fill}}p{0.1cm}cccc}
		\hline\hline
         & $R^x = R^y = R^z = 0$ & $R^z = 0$, $R^x$ or $R^y$ $\neq 0$ & $R^z \neq 0$, $R^x = R^y = 0$ & $R^z \neq 0$, $R^x$ or $R^y \neq 0$
          \\
		\hline
        1st$-$order
        &$\times$ & $\times$ &$\times$ &$\checkmark$
        \\
        \hline
	  2nd$-$order
	  & $\times$ &$\times$  &$\checkmark$ &$\checkmark$ \\
		\hline\hline
	\end{tabular*}
\end{table*}


\textit{{\color{blue}FBC-induced nonlinear PHE  in WSMs.-}}
We already know that when the tilting vector lies along the normal to the transport plane, the linear FBC-induced PHE vanishes and  the second-order response dominates. The nonlinear Hall effect in WSMs induced by the chiral anomaly as well as the Berry curvature dipole has been investigated in Refs  \cite{PhysRevB.103.045105, PhysRevB.97.041101, PhysRevB.103.245119, PhysRevLett.123.246602, PhysRevB.104.205124}. Here we show that the FBC can also induce a nonlinear PHE.  Recently, an intrinsic nonlinear PHE is proposed \cite{huang2022intrinsic}, where the effect of magnetic field plays a role via a magnetic-field-induced correction to the electric-field-induced correction of the Berry connection. For a comparison, we treat the effects of electric field and magnetic field on the same footing so that the proposed linear and nonlinear PHE are completely different from that in \cite{huang2022intrinsic}.

Combining Eqs. (\ref{distr-f1}), (\ref{distr-f2}) and (\ref{eq-mo-r}), the second-order electric current induced by FBC is obtained as
\begin{eqnarray}
\bm{J}^{(2)}_{\text{FBC}}&=&-\frac{e^3\tau}{\hbar}\int [d\bm{k}] \left[(\bm{\nabla}\times (\overleftrightarrow{\bm{G}}\bm{E}))\times (\bm{B}\times {\bm{v}}_{\bm{k}})(\bm{E}\cdot {\bm{v}}_{\bm{k}})\right.  \notag\\
&+& \left.  (\bm{E}\cdot \bm{B}){\bm{v}}_{\bm{k}}(\bm{\nabla}\times (\overleftrightarrow{\bm{G}}\bm{E}))\cdot {\bm{v}}_{\bm{k}}\right. \notag\\
&+& \left. \bm{E}\times(\bm{\nabla}\times (\overleftrightarrow{\bm{F}}\bm{B}))\bm{E}\cdot {\bm{v}}_{\bm{k}}\right]\frac{\partial f_{0}}{\partial \varepsilon_{\bm{k}}},
\label{j2}
\end{eqnarray}
where the first two terms are contributed from the E-FBC, with a defined second-order PHC $\sigma_{\text{E-FBC}}^{\alpha\beta\gamma}$. The third term is the B-FBC contribution $\sigma_{\text{B-FBC}}^{\alpha\beta\gamma}$. The total second-order FBC-induced PHC is then  $\sigma_{\text{FBC}}^{yxx}=\sigma_{\text{E-FBC}}^{yxx}+\sigma_{\text{B-FBC}}^{yxx}$.

We then solve the second-order PHC tensors induced by E-FBC and B-FBC, which are given by (the details are left to the Supplemental Material \cite{supp})
\begin{eqnarray}
\sigma^{yxx}_{\text{E-FBC}}&=&-\frac{e^3\tau}{\hbar}\sum_{s}\int [d\bm{k}]\frac{\partial f_{0}}{\partial \varepsilon_{\bm{k}}}\frac{1}{2v_F k^5}\left[v_F k^y R^{z}_{s}v^{y}_{s} B^x\right. \notag \\
 &-& \left. v_Fk^z R^{y}_{s}v^{y}_{s} B^x +k^y v^{z}_{s} v^{x}_{s} B^y  \right], \label{sig-yxx-e} \\
\sigma^{yxx}_{\text{B-FBC}}&=& -\frac{e^3\tau}{\hbar}\sum_{s}\int [d\bm{k}]\frac{\partial f_{0}}{\partial \varepsilon_{\bm{k}}}\frac{v^x}{2k^6}\left[k^y (2k^z + k R^{z}_{s})B^y \right. \notag \\
 &+& \left. k^x (2k^z + k R^{z}_{s})B^x \right].\label{sig-yxx-b}
\end{eqnarray}

We consider that a pair of Weyl cones are tilted in the same direction with $\bm{R}_{s}=\bm{R}$. 
One observes that a finite value of the first term in $\sigma^{yxx}_{\text{E-FBC}}$ requires a nonzero $R^z$. For the second term in $\sigma^{yxx}_{\text{E-FBC}}$, it is proportional to $k^{z}(k^y /k + R_{s}^{y})$ that takes a finite value if the tilting vector has both nonzero components along $y$ and $z$ axis.
Similarly, the third term in $\sigma^{yxx}_{\text{E-FBC}}$ takes a finite value if the $x$, $y$ and $z$ components of the tilting are nonzero simultaneously.
Similar deduction applies to $\sigma^{yxx}_{\text{B-FBC}}$.  Recall that a nonvanishing linear FBC-induced PHE requires that the angle ($\alpha$) between the tilt and the normal of transport plane is not zero, nor $\pi/2$. Thus, when $R^z$ is finite and  $R^x=R^y=0$, the linear PHE vanishes and the second-order PHE dominates, with a second-order PHC  \cite{supp}
\begin{equation}\label{sig-sec-f}
\begin{aligned}
\sigma^{yxx}_{\text{FBC}}=-\frac{e^3\tau}{\hbar}v_{F}\int [d\bm{k}]&\frac{\partial f_{0}}{\partial \varepsilon_{\bm{k}}}\frac{(k^x)^2 +(k^y)^2}{k^6}R^z B^x.
\end{aligned}
\end{equation}
The constrains on the linear and second-order PHCs from the tilting directions  are summarised in Table \ref{tab1}.

After some algebra, the analytical expression of \Eq{sig-sec-f} is found as \cite{supp}
\begin{equation}\label{sig-bcp-2-para}
\begin{aligned}
\sigma^{yxx}_{\text{FBC}} =-\frac{e^3\tau v_{F}^3\pi}{15\hbar\mu^2}\frac{ [5+(R^{z})^2]R^z}{\sqrt{1-(R^{z})^{2}}}B^x.
\end{aligned}
\end{equation}
As seen in \Eq{sig-bcp-2-para}  the nonlinear PHC scales with $\mu^{-2}$, which indicates that the nonlinear PHC increases rapidly when the Fermi level approaches to the Weyl nodes. This characteristic is due to the singularity of the quantum metric at the nodes.
 The FBC-induced nonlinear PHE predicted here vanishes when the tiling vector lies in the transport plane, while the chiral-anomaly-induced nonlinear PHE survives \cite{PhysRevB.103.045105, PhysRevB.104.205124}.


\textit{{\color{blue}Conclusions and Discussions.-}}
In this work we propose a new type of  PHE  originating from the field-induced Berry connection in tilted WSMs, which is closely associated with the quantum metric. As a  novel transport phenomenon, it is distinct from  the most remarkable transport phenomena for the WSMs, e.g., the negative longitudinal magnetoresistance and the extrinsic planar Hall effect, which are induced by the chiral anomaly. Due to the intrinsic nature, the PHE proposed here reflects the 
microscopic geometric properties of Bloch electrons, which could be useful for band structure engineering to the observation of the quantum metric by {\it{ab initio}} calculations. The FBC-induced nonlinear PHE is also studied. The conditions for the existence of the linear and nonlinear PHE are discussed, revealing that it is possible to distinguish them in experiments.


\begin{acknowledgments}
The authors thank Z. F. Zhang for helpful discussions.
This work is supported in part by the NSFC (Grants No. 11974348 and No. 11834014), and the National Key R\&D Program of China (Grant No.
2018YFA0305800). It is also supported by the Fundamental Research Funds for the Central Universities, and the Strategic Priority
Research Program of CAS (Grants No. XDB28000000, and No. XDB33000000). Z.G.Z. is supported in part by the Training Program of Major Research plan of
the National Natural Science Foundation of China (Grant No. 92165105).
\end{acknowledgments}


\bibliographystyle{apsrev4-2}

\begin{thebibliography}{61}%
\makeatletter
\providecommand \@ifxundefined [1]{%
 \@ifx{#1\undefined}
}%
\providecommand \@ifnum [1]{%
 \ifnum #1\expandafter \@firstoftwo
 \else \expandafter \@secondoftwo
 \fi
}%
\providecommand \@ifx [1]{%
 \ifx #1\expandafter \@firstoftwo
 \else \expandafter \@secondoftwo
 \fi
}%
\providecommand \natexlab [1]{#1}%
\providecommand \enquote  [1]{``#1''}%
\providecommand \bibnamefont  [1]{#1}%
\providecommand \bibfnamefont [1]{#1}%
\providecommand \citenamefont [1]{#1}%
\providecommand \href@noop [0]{\@secondoftwo}%
\providecommand \href [0]{\begingroup \@sanitize@url \@href}%
\providecommand \@href[1]{\@@startlink{#1}\@@href}%
\providecommand \@@href[1]{\endgroup#1\@@endlink}%
\providecommand \@sanitize@url [0]{\catcode `\\12\catcode `\$12\catcode
  `\&12\catcode `\#12\catcode `\^12\catcode `\_12\catcode `\%12\relax}%
\providecommand \@@startlink[1]{}%
\providecommand \@@endlink[0]{}%
\providecommand \url  [0]{\begingroup\@sanitize@url \@url }%
\providecommand \@url [1]{\endgroup\@href {#1}{\urlprefix }}%
\providecommand \urlprefix  [0]{URL }%
\providecommand \Eprint [0]{\href }%
\providecommand \doibase [0]{https://doi.org/}%
\providecommand \selectlanguage [0]{\@gobble}%
\providecommand \bibinfo  [0]{\@secondoftwo}%
\providecommand \bibfield  [0]{\@secondoftwo}%
\providecommand \translation [1]{[#1]}%
\providecommand \BibitemOpen [0]{}%
\providecommand \bibitemStop [0]{}%
\providecommand \bibitemNoStop [0]{.\EOS\space}%
\providecommand \EOS [0]{\spacefactor3000\relax}%
\providecommand \BibitemShut  [1]{\csname bibitem#1\endcsname}%
\let\auto@bib@innerbib\@empty
\bibitem [{\citenamefont {Wan}\ \emph {et~al.}(2011)\citenamefont {Wan},
  \citenamefont {Turner}, \citenamefont {Vishwanath},\ and\ \citenamefont
  {Savrasov}}]{PhysRevB.83.205101}%
  \BibitemOpen
  \bibfield  {author} {\bibinfo {author} {\bibfnamefont {X.}~\bibnamefont
  {Wan}}, \bibinfo {author} {\bibfnamefont {A.~M.}\ \bibnamefont {Turner}},
  \bibinfo {author} {\bibfnamefont {A.}~\bibnamefont {Vishwanath}},\ and\
  \bibinfo {author} {\bibfnamefont {S.~Y.}\ \bibnamefont {Savrasov}},\ }\href
  {https://doi.org/10.1103/PhysRevB.83.205101} {\bibfield  {journal} {\bibinfo
  {journal} {Phys. Rev. B}\ }\textbf {\bibinfo {volume} {83}},\ \bibinfo
  {pages} {205101} (\bibinfo {year} {2011})}\BibitemShut {NoStop}%
\bibitem [{\citenamefont {Weng}\ \emph {et~al.}(2015)\citenamefont {Weng},
  \citenamefont {Fang}, \citenamefont {Fang}, \citenamefont {Bernevig},\ and\
  \citenamefont {Dai}}]{PhysRevX.5.011029}%
  \BibitemOpen
  \bibfield  {author} {\bibinfo {author} {\bibfnamefont {H.}~\bibnamefont
  {Weng}}, \bibinfo {author} {\bibfnamefont {C.}~\bibnamefont {Fang}}, \bibinfo
  {author} {\bibfnamefont {Z.}~\bibnamefont {Fang}}, \bibinfo {author}
  {\bibfnamefont {B.~A.}\ \bibnamefont {Bernevig}},\ and\ \bibinfo {author}
  {\bibfnamefont {X.}~\bibnamefont {Dai}},\ }\href
  {https://doi.org/10.1103/PhysRevX.5.011029} {\bibfield  {journal} {\bibinfo
  {journal} {Phys. Rev. X}\ }\textbf {\bibinfo {volume} {5}},\ \bibinfo {pages}
  {011029} (\bibinfo {year} {2015})}\BibitemShut {NoStop}%
\bibitem [{\citenamefont {Lv}\ \emph {et~al.}(2015)\citenamefont {Lv},
  \citenamefont {Weng}, \citenamefont {Fu}, \citenamefont {Wang}, \citenamefont
  {Miao}, \citenamefont {Ma}, \citenamefont {Richard}, \citenamefont {Huang},
  \citenamefont {Zhao}, \citenamefont {Chen}, \citenamefont {Fang},
  \citenamefont {Dai}, \citenamefont {Qian},\ and\ \citenamefont
  {Ding}}]{PhysRevX.5.031013}%
  \BibitemOpen
  \bibfield  {author} {\bibinfo {author} {\bibfnamefont {B.~Q.}\ \bibnamefont
  {Lv}}, \bibinfo {author} {\bibfnamefont {H.~M.}\ \bibnamefont {Weng}},
  \bibinfo {author} {\bibfnamefont {B.~B.}\ \bibnamefont {Fu}}, \bibinfo
  {author} {\bibfnamefont {X.~P.}\ \bibnamefont {Wang}}, \bibinfo {author}
  {\bibfnamefont {H.}~\bibnamefont {Miao}}, \bibinfo {author} {\bibfnamefont
  {J.}~\bibnamefont {Ma}}, \bibinfo {author} {\bibfnamefont {P.}~\bibnamefont
  {Richard}}, \bibinfo {author} {\bibfnamefont {X.~C.}\ \bibnamefont {Huang}},
  \bibinfo {author} {\bibfnamefont {L.~X.}\ \bibnamefont {Zhao}}, \bibinfo
  {author} {\bibfnamefont {G.~F.}\ \bibnamefont {Chen}}, \bibinfo {author}
  {\bibfnamefont {Z.}~\bibnamefont {Fang}}, \bibinfo {author} {\bibfnamefont
  {X.}~\bibnamefont {Dai}}, \bibinfo {author} {\bibfnamefont {T.}~\bibnamefont
  {Qian}},\ and\ \bibinfo {author} {\bibfnamefont {H.}~\bibnamefont {Ding}},\
  }\href {https://doi.org/10.1103/PhysRevX.5.031013} {\bibfield  {journal}
  {\bibinfo  {journal} {Phys. Rev. X}\ }\textbf {\bibinfo {volume} {5}},\
  \bibinfo {pages} {031013} (\bibinfo {year} {2015})}\BibitemShut {NoStop}%
\bibitem [{\citenamefont {Xu}\ \emph {et~al.}(2015{\natexlab{a}})\citenamefont
  {Xu}, \citenamefont {Belopolski}, \citenamefont {Alidoust}, \citenamefont
  {Neupane}, \citenamefont {Bian}, \citenamefont {Zhang}, \citenamefont
  {Sankar}, \citenamefont {Chang}, \citenamefont {Yuan}, \citenamefont {Lee},
  \citenamefont {Huang}, \citenamefont {Zheng}, \citenamefont {Ma},
  \citenamefont {Sanchez}, \citenamefont {Wang}, \citenamefont {Bansil},
  \citenamefont {Chou}, \citenamefont {Shibayev}, \citenamefont {Lin},
  \citenamefont {Jia},\ and\ \citenamefont {Hasan}}]{xu2015}%
  \BibitemOpen
  \bibfield  {author} {\bibinfo {author} {\bibfnamefont {S.-Y.}\ \bibnamefont
  {Xu}}, \bibinfo {author} {\bibfnamefont {I.}~\bibnamefont {Belopolski}},
  \bibinfo {author} {\bibfnamefont {N.}~\bibnamefont {Alidoust}}, \bibinfo
  {author} {\bibfnamefont {M.}~\bibnamefont {Neupane}}, \bibinfo {author}
  {\bibfnamefont {G.}~\bibnamefont {Bian}}, \bibinfo {author} {\bibfnamefont
  {C.}~\bibnamefont {Zhang}}, \bibinfo {author} {\bibfnamefont
  {R.}~\bibnamefont {Sankar}}, \bibinfo {author} {\bibfnamefont
  {G.}~\bibnamefont {Chang}}, \bibinfo {author} {\bibfnamefont
  {Z.}~\bibnamefont {Yuan}}, \bibinfo {author} {\bibfnamefont {C.-C.}\
  \bibnamefont {Lee}}, \bibinfo {author} {\bibfnamefont {S.-M.}\ \bibnamefont
  {Huang}}, \bibinfo {author} {\bibfnamefont {H.}~\bibnamefont {Zheng}},
  \bibinfo {author} {\bibfnamefont {J.}~\bibnamefont {Ma}}, \bibinfo {author}
  {\bibfnamefont {D.~S.}\ \bibnamefont {Sanchez}}, \bibinfo {author}
  {\bibfnamefont {B.}~\bibnamefont {Wang}}, \bibinfo {author} {\bibfnamefont
  {A.}~\bibnamefont {Bansil}}, \bibinfo {author} {\bibfnamefont
  {F.}~\bibnamefont {Chou}}, \bibinfo {author} {\bibfnamefont {P.~P.}\
  \bibnamefont {Shibayev}}, \bibinfo {author} {\bibfnamefont {H.}~\bibnamefont
  {Lin}}, \bibinfo {author} {\bibfnamefont {S.}~\bibnamefont {Jia}},\ and\
  \bibinfo {author} {\bibfnamefont {M.~Z.}\ \bibnamefont {Hasan}},\ }\href
  {https://doi.org/10.1126/science.aaa9297} {\bibfield  {journal} {\bibinfo
  {journal} {Science}\ }\textbf {\bibinfo {volume} {349}},\ \bibinfo {pages}
  {613} (\bibinfo {year} {2015}{\natexlab{a}})},\ \Eprint
  {https://arxiv.org/abs/https://www.science.org/doi/pdf/10.1126/science.aaa9297}
  {https://www.science.org/doi/pdf/10.1126/science.aaa9297} \BibitemShut
  {NoStop}%
\bibitem [{\citenamefont {Xu}\ \emph {et~al.}(2015{\natexlab{b}})\citenamefont
  {Xu}, \citenamefont {Belopolski}, \citenamefont {Sanchez}, \citenamefont
  {Zhang}, \citenamefont {Chang}, \citenamefont {Guo}, \citenamefont {Bian},
  \citenamefont {Yuan}, \citenamefont {Lu}, \citenamefont {Chang},
  \citenamefont {Shibayev}, \citenamefont {Prokopovych}, \citenamefont
  {Alidoust}, \citenamefont {Zheng}, \citenamefont {Lee}, \citenamefont
  {Huang}, \citenamefont {Sankar}, \citenamefont {Chou}, \citenamefont {Hsu},
  \citenamefont {Jeng}, \citenamefont {Bansil}, \citenamefont {Neupert},
  \citenamefont {Strocov}, \citenamefont {Lin}, \citenamefont {Jia},\ and\
  \citenamefont {Hasan}}]{xu2015-2}%
  \BibitemOpen
  \bibfield  {author} {\bibinfo {author} {\bibfnamefont {S.-Y.}\ \bibnamefont
  {Xu}}, \bibinfo {author} {\bibfnamefont {I.}~\bibnamefont {Belopolski}},
  \bibinfo {author} {\bibfnamefont {D.~S.}\ \bibnamefont {Sanchez}}, \bibinfo
  {author} {\bibfnamefont {C.}~\bibnamefont {Zhang}}, \bibinfo {author}
  {\bibfnamefont {G.}~\bibnamefont {Chang}}, \bibinfo {author} {\bibfnamefont
  {C.}~\bibnamefont {Guo}}, \bibinfo {author} {\bibfnamefont {G.}~\bibnamefont
  {Bian}}, \bibinfo {author} {\bibfnamefont {Z.}~\bibnamefont {Yuan}}, \bibinfo
  {author} {\bibfnamefont {H.}~\bibnamefont {Lu}}, \bibinfo {author}
  {\bibfnamefont {T.-R.}\ \bibnamefont {Chang}}, \bibinfo {author}
  {\bibfnamefont {P.~P.}\ \bibnamefont {Shibayev}}, \bibinfo {author}
  {\bibfnamefont {M.~L.}\ \bibnamefont {Prokopovych}}, \bibinfo {author}
  {\bibfnamefont {N.}~\bibnamefont {Alidoust}}, \bibinfo {author}
  {\bibfnamefont {H.}~\bibnamefont {Zheng}}, \bibinfo {author} {\bibfnamefont
  {C.-C.}\ \bibnamefont {Lee}}, \bibinfo {author} {\bibfnamefont {S.-M.}\
  \bibnamefont {Huang}}, \bibinfo {author} {\bibfnamefont {R.}~\bibnamefont
  {Sankar}}, \bibinfo {author} {\bibfnamefont {F.}~\bibnamefont {Chou}},
  \bibinfo {author} {\bibfnamefont {C.-H.}\ \bibnamefont {Hsu}}, \bibinfo
  {author} {\bibfnamefont {H.-T.}\ \bibnamefont {Jeng}}, \bibinfo {author}
  {\bibfnamefont {A.}~\bibnamefont {Bansil}}, \bibinfo {author} {\bibfnamefont
  {T.}~\bibnamefont {Neupert}}, \bibinfo {author} {\bibfnamefont {V.~N.}\
  \bibnamefont {Strocov}}, \bibinfo {author} {\bibfnamefont {H.}~\bibnamefont
  {Lin}}, \bibinfo {author} {\bibfnamefont {S.}~\bibnamefont {Jia}},\ and\
  \bibinfo {author} {\bibfnamefont {M.~Z.}\ \bibnamefont {Hasan}},\ }\href
  {https://doi.org/10.1126/sciadv.1501092} {\bibfield  {journal} {\bibinfo
  {journal} {Science Advances}\ }\textbf {\bibinfo {volume} {1}},\ \bibinfo
  {pages} {e1501092} (\bibinfo {year} {2015}{\natexlab{b}})},\ \Eprint
  {https://arxiv.org/abs/https://www.science.org/doi/pdf/10.1126/sciadv.1501092}
  {https://www.science.org/doi/pdf/10.1126/sciadv.1501092} \BibitemShut
  {NoStop}%
\bibitem [{\citenamefont {Hasan}\ \emph {et~al.}(2017)\citenamefont {Hasan},
  \citenamefont {Xu}, \citenamefont {Belopolski},\ and\ \citenamefont
  {Huang}}]{doi:10.1146/annurev-conmatphys-031016-025225}%
  \BibitemOpen
  \bibfield  {author} {\bibinfo {author} {\bibfnamefont {M.~Z.}\ \bibnamefont
  {Hasan}}, \bibinfo {author} {\bibfnamefont {S.-Y.}\ \bibnamefont {Xu}},
  \bibinfo {author} {\bibfnamefont {I.}~\bibnamefont {Belopolski}},\ and\
  \bibinfo {author} {\bibfnamefont {S.-M.}\ \bibnamefont {Huang}},\ }\href
  {https://doi.org/10.1146/annurev-conmatphys-031016-025225} {\bibfield
  {journal} {\bibinfo  {journal} {Annual Review of Condensed Matter Physics}\
  }\textbf {\bibinfo {volume} {8}},\ \bibinfo {pages} {289} (\bibinfo {year}
  {2017})},\ \Eprint
  {https://arxiv.org/abs/https://doi.org/10.1146/annurev-conmatphys-031016-025225}
  {https://doi.org/10.1146/annurev-conmatphys-031016-025225} \BibitemShut
  {NoStop}%
\bibitem [{\citenamefont {Yan}\ and\ \citenamefont
  {Felser}(2017)}]{doi:10.1146/annurev-conmatphys-031016-025458}%
  \BibitemOpen
  \bibfield  {author} {\bibinfo {author} {\bibfnamefont {B.}~\bibnamefont
  {Yan}}\ and\ \bibinfo {author} {\bibfnamefont {C.}~\bibnamefont {Felser}},\
  }\href {https://doi.org/10.1146/annurev-conmatphys-031016-025458} {\bibfield
  {journal} {\bibinfo  {journal} {Annual Review of Condensed Matter Physics}\
  }\textbf {\bibinfo {volume} {8}},\ \bibinfo {pages} {337} (\bibinfo {year}
  {2017})},\ \Eprint
  {https://arxiv.org/abs/https://doi.org/10.1146/annurev-conmatphys-031016-025458}
  {https://doi.org/10.1146/annurev-conmatphys-031016-025458} \BibitemShut
  {NoStop}%
\bibitem [{\citenamefont {Armitage}\ \emph {et~al.}(2018)\citenamefont
  {Armitage}, \citenamefont {Mele},\ and\ \citenamefont
  {Vishwanath}}]{RevModPhys.90.015001}%
  \BibitemOpen
  \bibfield  {author} {\bibinfo {author} {\bibfnamefont {N.~P.}\ \bibnamefont
  {Armitage}}, \bibinfo {author} {\bibfnamefont {E.~J.}\ \bibnamefont {Mele}},\
  and\ \bibinfo {author} {\bibfnamefont {A.}~\bibnamefont {Vishwanath}},\
  }\href {https://doi.org/10.1103/RevModPhys.90.015001} {\bibfield  {journal}
  {\bibinfo  {journal} {Rev. Mod. Phys.}\ }\textbf {\bibinfo {volume} {90}},\
  \bibinfo {pages} {015001} (\bibinfo {year} {2018})}\BibitemShut {NoStop}%
\bibitem [{\citenamefont {Yang}\ \emph {et~al.}(2011)\citenamefont {Yang},
  \citenamefont {Lu},\ and\ \citenamefont {Ran}}]{PhysRevB.84.075129}%
  \BibitemOpen
  \bibfield  {author} {\bibinfo {author} {\bibfnamefont {K.-Y.}\ \bibnamefont
  {Yang}}, \bibinfo {author} {\bibfnamefont {Y.-M.}\ \bibnamefont {Lu}},\ and\
  \bibinfo {author} {\bibfnamefont {Y.}~\bibnamefont {Ran}},\ }\href
  {https://doi.org/10.1103/PhysRevB.84.075129} {\bibfield  {journal} {\bibinfo
  {journal} {Phys. Rev. B}\ }\textbf {\bibinfo {volume} {84}},\ \bibinfo
  {pages} {075129} (\bibinfo {year} {2011})}\BibitemShut {NoStop}%
\bibitem [{\citenamefont {Burkov}\ and\ \citenamefont
  {Balents}(2011)}]{PhysRevLett.107.127205}%
  \BibitemOpen
  \bibfield  {author} {\bibinfo {author} {\bibfnamefont {A.~A.}\ \bibnamefont
  {Burkov}}\ and\ \bibinfo {author} {\bibfnamefont {L.}~\bibnamefont
  {Balents}},\ }\href {https://doi.org/10.1103/PhysRevLett.107.127205}
  {\bibfield  {journal} {\bibinfo  {journal} {Phys. Rev. Lett.}\ }\textbf
  {\bibinfo {volume} {107}},\ \bibinfo {pages} {127205} (\bibinfo {year}
  {2011})}\BibitemShut {NoStop}%
\bibitem [{\citenamefont {Xu}\ \emph {et~al.}(2011)\citenamefont {Xu},
  \citenamefont {Weng}, \citenamefont {Wang}, \citenamefont {Dai},\ and\
  \citenamefont {Fang}}]{PhysRevLett.107.186806}%
  \BibitemOpen
  \bibfield  {author} {\bibinfo {author} {\bibfnamefont {G.}~\bibnamefont
  {Xu}}, \bibinfo {author} {\bibfnamefont {H.}~\bibnamefont {Weng}}, \bibinfo
  {author} {\bibfnamefont {Z.}~\bibnamefont {Wang}}, \bibinfo {author}
  {\bibfnamefont {X.}~\bibnamefont {Dai}},\ and\ \bibinfo {author}
  {\bibfnamefont {Z.}~\bibnamefont {Fang}},\ }\href
  {https://doi.org/10.1103/PhysRevLett.107.186806} {\bibfield  {journal}
  {\bibinfo  {journal} {Phys. Rev. Lett.}\ }\textbf {\bibinfo {volume} {107}},\
  \bibinfo {pages} {186806} (\bibinfo {year} {2011})}\BibitemShut {NoStop}%
\bibitem [{\citenamefont {Jiang}\ \emph {et~al.}(2021)\citenamefont {Jiang},
  \citenamefont {de~Sousa}, \citenamefont {Wang},\ and\ \citenamefont
  {Low}}]{PhysRevLett.126.106601}%
  \BibitemOpen
  \bibfield  {author} {\bibinfo {author} {\bibfnamefont {W.}~\bibnamefont
  {Jiang}}, \bibinfo {author} {\bibfnamefont {D.~J.~P.}\ \bibnamefont
  {de~Sousa}}, \bibinfo {author} {\bibfnamefont {J.-P.}\ \bibnamefont {Wang}},\
  and\ \bibinfo {author} {\bibfnamefont {T.}~\bibnamefont {Low}},\ }\href
  {https://doi.org/10.1103/PhysRevLett.126.106601} {\bibfield  {journal}
  {\bibinfo  {journal} {Phys. Rev. Lett.}\ }\textbf {\bibinfo {volume} {126}},\
  \bibinfo {pages} {106601} (\bibinfo {year} {2021})}\BibitemShut {NoStop}%
\bibitem [{\citenamefont {Son}\ and\ \citenamefont
  {Spivak}(2013)}]{PhysRevB.88.104412}%
  \BibitemOpen
  \bibfield  {author} {\bibinfo {author} {\bibfnamefont {D.~T.}\ \bibnamefont
  {Son}}\ and\ \bibinfo {author} {\bibfnamefont {B.~Z.}\ \bibnamefont
  {Spivak}},\ }\href {https://doi.org/10.1103/PhysRevB.88.104412} {\bibfield
  {journal} {\bibinfo  {journal} {Phys. Rev. B}\ }\textbf {\bibinfo {volume}
  {88}},\ \bibinfo {pages} {104412} (\bibinfo {year} {2013})}\BibitemShut
  {NoStop}%
\bibitem [{\citenamefont {Burkov}(2014)}]{PhysRevLett.113.247203}%
  \BibitemOpen
  \bibfield  {author} {\bibinfo {author} {\bibfnamefont {A.~A.}\ \bibnamefont
  {Burkov}},\ }\href {https://doi.org/10.1103/PhysRevLett.113.247203}
  {\bibfield  {journal} {\bibinfo  {journal} {Phys. Rev. Lett.}\ }\textbf
  {\bibinfo {volume} {113}},\ \bibinfo {pages} {247203} (\bibinfo {year}
  {2014})}\BibitemShut {NoStop}%
\bibitem [{\citenamefont {Kharzeev}\ and\ \citenamefont
  {Yee}(2013)}]{PhysRevB.88.115119}%
  \BibitemOpen
  \bibfield  {author} {\bibinfo {author} {\bibfnamefont {D.~E.}\ \bibnamefont
  {Kharzeev}}\ and\ \bibinfo {author} {\bibfnamefont {H.-U.}\ \bibnamefont
  {Yee}},\ }\href {https://doi.org/10.1103/PhysRevB.88.115119} {\bibfield
  {journal} {\bibinfo  {journal} {Phys. Rev. B}\ }\textbf {\bibinfo {volume}
  {88}},\ \bibinfo {pages} {115119} (\bibinfo {year} {2013})}\BibitemShut
  {NoStop}%
\bibitem [{\citenamefont {Parameswaran}\ \emph {et~al.}(2014)\citenamefont
  {Parameswaran}, \citenamefont {Grover}, \citenamefont {Abanin}, \citenamefont
  {Pesin},\ and\ \citenamefont {Vishwanath}}]{PhysRevX.4.031035}%
  \BibitemOpen
  \bibfield  {author} {\bibinfo {author} {\bibfnamefont {S.~A.}\ \bibnamefont
  {Parameswaran}}, \bibinfo {author} {\bibfnamefont {T.}~\bibnamefont
  {Grover}}, \bibinfo {author} {\bibfnamefont {D.~A.}\ \bibnamefont {Abanin}},
  \bibinfo {author} {\bibfnamefont {D.~A.}\ \bibnamefont {Pesin}},\ and\
  \bibinfo {author} {\bibfnamefont {A.}~\bibnamefont {Vishwanath}},\ }\href
  {https://doi.org/10.1103/PhysRevX.4.031035} {\bibfield  {journal} {\bibinfo
  {journal} {Phys. Rev. X}\ }\textbf {\bibinfo {volume} {4}},\ \bibinfo {pages}
  {031035} (\bibinfo {year} {2014})}\BibitemShut {NoStop}%
\bibitem [{\citenamefont {Burkov}(2015)}]{Burkov_2015}%
  \BibitemOpen
  \bibfield  {author} {\bibinfo {author} {\bibfnamefont {A.~A.}\ \bibnamefont
  {Burkov}},\ }\href {https://doi.org/10.1088/0953-8984/27/11/113201}
  {\bibfield  {journal} {\bibinfo  {journal} {Journal of Physics: Condensed
  Matter}\ }\textbf {\bibinfo {volume} {27}},\ \bibinfo {pages} {113201}
  (\bibinfo {year} {2015})}\BibitemShut {NoStop}%
\bibitem [{\citenamefont {Zyuzin}(2017)}]{PhysRevB.95.245128}%
  \BibitemOpen
  \bibfield  {author} {\bibinfo {author} {\bibfnamefont {V.~A.}\ \bibnamefont
  {Zyuzin}},\ }\href {https://doi.org/10.1103/PhysRevB.95.245128} {\bibfield
  {journal} {\bibinfo  {journal} {Phys. Rev. B}\ }\textbf {\bibinfo {volume}
  {95}},\ \bibinfo {pages} {245128} (\bibinfo {year} {2017})}\BibitemShut
  {NoStop}%
\bibitem [{\citenamefont {Kim}\ \emph {et~al.}(2014)\citenamefont {Kim},
  \citenamefont {Kim},\ and\ \citenamefont {Sasaki}}]{PhysRevB.89.195137}%
  \BibitemOpen
  \bibfield  {author} {\bibinfo {author} {\bibfnamefont {K.-S.}\ \bibnamefont
  {Kim}}, \bibinfo {author} {\bibfnamefont {H.-J.}\ \bibnamefont {Kim}},\ and\
  \bibinfo {author} {\bibfnamefont {M.}~\bibnamefont {Sasaki}},\ }\href
  {https://doi.org/10.1103/PhysRevB.89.195137} {\bibfield  {journal} {\bibinfo
  {journal} {Phys. Rev. B}\ }\textbf {\bibinfo {volume} {89}},\ \bibinfo
  {pages} {195137} (\bibinfo {year} {2014})}\BibitemShut {NoStop}%
\bibitem [{\citenamefont {Nandy}\ \emph {et~al.}(2017)\citenamefont {Nandy},
  \citenamefont {Sharma}, \citenamefont {Taraphder},\ and\ \citenamefont
  {Tewari}}]{PhysRevLett.119.176804}%
  \BibitemOpen
  \bibfield  {author} {\bibinfo {author} {\bibfnamefont {S.}~\bibnamefont
  {Nandy}}, \bibinfo {author} {\bibfnamefont {G.}~\bibnamefont {Sharma}},
  \bibinfo {author} {\bibfnamefont {A.}~\bibnamefont {Taraphder}},\ and\
  \bibinfo {author} {\bibfnamefont {S.}~\bibnamefont {Tewari}},\ }\href
  {https://doi.org/10.1103/PhysRevLett.119.176804} {\bibfield  {journal}
  {\bibinfo  {journal} {Phys. Rev. Lett.}\ }\textbf {\bibinfo {volume} {119}},\
  \bibinfo {pages} {176804} (\bibinfo {year} {2017})}\BibitemShut {NoStop}%
\bibitem [{\citenamefont {Burkov}(2017)}]{PhysRevB.96.041110}%
  \BibitemOpen
  \bibfield  {author} {\bibinfo {author} {\bibfnamefont {A.~A.}\ \bibnamefont
  {Burkov}},\ }\href {https://doi.org/10.1103/PhysRevB.96.041110} {\bibfield
  {journal} {\bibinfo  {journal} {Phys. Rev. B}\ }\textbf {\bibinfo {volume}
  {96}},\ \bibinfo {pages} {041110} (\bibinfo {year} {2017})}\BibitemShut
  {NoStop}%
\bibitem [{\citenamefont {Ghosh}\ \emph {et~al.}(2020)\citenamefont {Ghosh},
  \citenamefont {Sinha}, \citenamefont {Nandy},\ and\ \citenamefont
  {Taraphder}}]{PhysRevB.102.121105}%
  \BibitemOpen
  \bibfield  {author} {\bibinfo {author} {\bibfnamefont {S.}~\bibnamefont
  {Ghosh}}, \bibinfo {author} {\bibfnamefont {D.}~\bibnamefont {Sinha}},
  \bibinfo {author} {\bibfnamefont {S.}~\bibnamefont {Nandy}},\ and\ \bibinfo
  {author} {\bibfnamefont {A.}~\bibnamefont {Taraphder}},\ }\href
  {https://doi.org/10.1103/PhysRevB.102.121105} {\bibfield  {journal} {\bibinfo
   {journal} {Phys. Rev. B}\ }\textbf {\bibinfo {volume} {102}},\ \bibinfo
  {pages} {121105} (\bibinfo {year} {2020})}\BibitemShut {NoStop}%
\bibitem [{\citenamefont {Son}\ and\ \citenamefont
  {Yamamoto}(2012)}]{PhysRevLett.109.181602}%
  \BibitemOpen
  \bibfield  {author} {\bibinfo {author} {\bibfnamefont {D.~T.}\ \bibnamefont
  {Son}}\ and\ \bibinfo {author} {\bibfnamefont {N.}~\bibnamefont {Yamamoto}},\
  }\href {https://doi.org/10.1103/PhysRevLett.109.181602} {\bibfield  {journal}
  {\bibinfo  {journal} {Phys. Rev. Lett.}\ }\textbf {\bibinfo {volume} {109}},\
  \bibinfo {pages} {181602} (\bibinfo {year} {2012})}\BibitemShut {NoStop}%
\bibitem [{\citenamefont {Stephanov}\ and\ \citenamefont
  {Yin}(2012)}]{PhysRevLett.109.162001}%
  \BibitemOpen
  \bibfield  {author} {\bibinfo {author} {\bibfnamefont {M.~A.}\ \bibnamefont
  {Stephanov}}\ and\ \bibinfo {author} {\bibfnamefont {Y.}~\bibnamefont
  {Yin}},\ }\href {https://doi.org/10.1103/PhysRevLett.109.162001} {\bibfield
  {journal} {\bibinfo  {journal} {Phys. Rev. Lett.}\ }\textbf {\bibinfo
  {volume} {109}},\ \bibinfo {pages} {162001} (\bibinfo {year}
  {2012})}\BibitemShut {NoStop}%
\bibitem [{\citenamefont {Landsteiner}\ \emph {et~al.}(2011)\citenamefont
  {Landsteiner}, \citenamefont {Meg\'{\i}as},\ and\ \citenamefont
  {Pena-Benitez}}]{PhysRevLett.107.021601}%
  \BibitemOpen
  \bibfield  {author} {\bibinfo {author} {\bibfnamefont {K.}~\bibnamefont
  {Landsteiner}}, \bibinfo {author} {\bibfnamefont {E.}~\bibnamefont
  {Meg\'{\i}as}},\ and\ \bibinfo {author} {\bibfnamefont {F.}~\bibnamefont
  {Pena-Benitez}},\ }\href {https://doi.org/10.1103/PhysRevLett.107.021601}
  {\bibfield  {journal} {\bibinfo  {journal} {Phys. Rev. Lett.}\ }\textbf
  {\bibinfo {volume} {107}},\ \bibinfo {pages} {021601} (\bibinfo {year}
  {2011})}\BibitemShut {NoStop}%
\bibitem [{\citenamefont {Chang}\ and\ \citenamefont
  {Yang}(2015)}]{PhysRevB.91.115203}%
  \BibitemOpen
  \bibfield  {author} {\bibinfo {author} {\bibfnamefont {M.-C.}\ \bibnamefont
  {Chang}}\ and\ \bibinfo {author} {\bibfnamefont {M.-F.}\ \bibnamefont
  {Yang}},\ }\href {https://doi.org/10.1103/PhysRevB.91.115203} {\bibfield
  {journal} {\bibinfo  {journal} {Phys. Rev. B}\ }\textbf {\bibinfo {volume}
  {91}},\ \bibinfo {pages} {115203} (\bibinfo {year} {2015})}\BibitemShut
  {NoStop}%
\bibitem [{\citenamefont {Berry}(1984)}]{berry1984quantal}%
  \BibitemOpen
  \bibfield  {author} {\bibinfo {author} {\bibfnamefont {M.~V.}\ \bibnamefont
  {Berry}},\ }\href@noop {} {\bibfield  {journal} {\bibinfo  {journal}
  {Proceedings of the Royal Society of London. A. Mathematical and Physical
  Sciences}\ }\textbf {\bibinfo {volume} {392}},\ \bibinfo {pages} {45}
  (\bibinfo {year} {1984})}\BibitemShut {NoStop}%
\bibitem [{\citenamefont {Berry}(1989)}]{berry1989geometric}%
  \BibitemOpen
  \bibfield  {author} {\bibinfo {author} {\bibfnamefont {M.}~\bibnamefont
  {Berry}},\ }\href@noop {} {\bibinfo {title} {Geometric phases in physics}}
  (\bibinfo {year} {1989})\BibitemShut {NoStop}%
\bibitem [{\citenamefont {Provost}\ and\ \citenamefont
  {Vallee}(1980)}]{provost1980riemannian}%
  \BibitemOpen
  \bibfield  {author} {\bibinfo {author} {\bibfnamefont {J.}~\bibnamefont
  {Provost}}\ and\ \bibinfo {author} {\bibfnamefont {G.}~\bibnamefont
  {Vallee}},\ }\href@noop {} {\bibfield  {journal} {\bibinfo  {journal}
  {Communications in Mathematical Physics}\ }\textbf {\bibinfo {volume} {76}},\
  \bibinfo {pages} {289} (\bibinfo {year} {1980})}\BibitemShut {NoStop}%
\bibitem [{\citenamefont {Bu\ifmmode~\check{z}\else \v{z}\fi{}ek}\ and\
  \citenamefont {Hillery}(1996)}]{PhysRevA.54.1844}%
  \BibitemOpen
  \bibfield  {author} {\bibinfo {author} {\bibfnamefont {V.}~\bibnamefont
  {Bu\ifmmode~\check{z}\else \v{z}\fi{}ek}}\ and\ \bibinfo {author}
  {\bibfnamefont {M.}~\bibnamefont {Hillery}},\ }\href
  {https://doi.org/10.1103/PhysRevA.54.1844} {\bibfield  {journal} {\bibinfo
  {journal} {Phys. Rev. A}\ }\textbf {\bibinfo {volume} {54}},\ \bibinfo
  {pages} {1844} (\bibinfo {year} {1996})}\BibitemShut {NoStop}%
\bibitem [{\citenamefont {Vedral}\ \emph {et~al.}(1997)\citenamefont {Vedral},
  \citenamefont {Plenio}, \citenamefont {Rippin},\ and\ \citenamefont
  {Knight}}]{PhysRevLett.78.2275}%
  \BibitemOpen
  \bibfield  {author} {\bibinfo {author} {\bibfnamefont {V.}~\bibnamefont
  {Vedral}}, \bibinfo {author} {\bibfnamefont {M.~B.}\ \bibnamefont {Plenio}},
  \bibinfo {author} {\bibfnamefont {M.~A.}\ \bibnamefont {Rippin}},\ and\
  \bibinfo {author} {\bibfnamefont {P.~L.}\ \bibnamefont {Knight}},\ }\href
  {https://doi.org/10.1103/PhysRevLett.78.2275} {\bibfield  {journal} {\bibinfo
   {journal} {Phys. Rev. Lett.}\ }\textbf {\bibinfo {volume} {78}},\ \bibinfo
  {pages} {2275} (\bibinfo {year} {1997})}\BibitemShut {NoStop}%
\bibitem [{\citenamefont {Baumgratz}\ \emph {et~al.}(2014)\citenamefont
  {Baumgratz}, \citenamefont {Cramer},\ and\ \citenamefont
  {Plenio}}]{PhysRevLett.113.140401}%
  \BibitemOpen
  \bibfield  {author} {\bibinfo {author} {\bibfnamefont {T.}~\bibnamefont
  {Baumgratz}}, \bibinfo {author} {\bibfnamefont {M.}~\bibnamefont {Cramer}},\
  and\ \bibinfo {author} {\bibfnamefont {M.~B.}\ \bibnamefont {Plenio}},\
  }\href {https://doi.org/10.1103/PhysRevLett.113.140401} {\bibfield  {journal}
  {\bibinfo  {journal} {Phys. Rev. Lett.}\ }\textbf {\bibinfo {volume} {113}},\
  \bibinfo {pages} {140401} (\bibinfo {year} {2014})}\BibitemShut {NoStop}%
\bibitem [{\citenamefont {\ifmmode~\dot{Z}\else \.{Z}\fi{}yczkowski}\ and\
  \citenamefont {Sommers}(2005)}]{PhysRevA.71.032313}%
  \BibitemOpen
  \bibfield  {author} {\bibinfo {author} {\bibfnamefont {K.}~\bibnamefont
  {\ifmmode~\dot{Z}\else \.{Z}\fi{}yczkowski}}\ and\ \bibinfo {author}
  {\bibfnamefont {H.-J.}\ \bibnamefont {Sommers}},\ }\href
  {https://doi.org/10.1103/PhysRevA.71.032313} {\bibfield  {journal} {\bibinfo
  {journal} {Phys. Rev. A}\ }\textbf {\bibinfo {volume} {71}},\ \bibinfo
  {pages} {032313} (\bibinfo {year} {2005})}\BibitemShut {NoStop}%
\bibitem [{\citenamefont {Leibfried}\ \emph {et~al.}(2003)\citenamefont
  {Leibfried}, \citenamefont {DeMarco}, \citenamefont {Meyer}, \citenamefont
  {Lucas}, \citenamefont {Barrett}, \citenamefont {Britton}, \citenamefont
  {Itano}, \citenamefont {Jelenkovi{\'c}}, \citenamefont {Langer},
  \citenamefont {Rosenband} \emph {et~al.}}]{leibfried2003experimental}%
  \BibitemOpen
  \bibfield  {author} {\bibinfo {author} {\bibfnamefont {D.}~\bibnamefont
  {Leibfried}}, \bibinfo {author} {\bibfnamefont {B.}~\bibnamefont {DeMarco}},
  \bibinfo {author} {\bibfnamefont {V.}~\bibnamefont {Meyer}}, \bibinfo
  {author} {\bibfnamefont {D.}~\bibnamefont {Lucas}}, \bibinfo {author}
  {\bibfnamefont {M.}~\bibnamefont {Barrett}}, \bibinfo {author} {\bibfnamefont
  {J.}~\bibnamefont {Britton}}, \bibinfo {author} {\bibfnamefont {W.~M.}\
  \bibnamefont {Itano}}, \bibinfo {author} {\bibfnamefont {B.}~\bibnamefont
  {Jelenkovi{\'c}}}, \bibinfo {author} {\bibfnamefont {C.}~\bibnamefont
  {Langer}}, \bibinfo {author} {\bibfnamefont {T.}~\bibnamefont {Rosenband}},
  \emph {et~al.},\ }\href@noop {} {\bibfield  {journal} {\bibinfo  {journal}
  {Nature}\ }\textbf {\bibinfo {volume} {422}},\ \bibinfo {pages} {412}
  (\bibinfo {year} {2003})}\BibitemShut {NoStop}%
\bibitem [{\citenamefont {Peotta}\ and\ \citenamefont
  {T{\"o}rm{\"a}}(2015)}]{peotta2015superfluidity}%
  \BibitemOpen
  \bibfield  {author} {\bibinfo {author} {\bibfnamefont {S.}~\bibnamefont
  {Peotta}}\ and\ \bibinfo {author} {\bibfnamefont {P.}~\bibnamefont
  {T{\"o}rm{\"a}}},\ }\href@noop {} {\bibfield  {journal} {\bibinfo  {journal}
  {Nature communications}\ }\textbf {\bibinfo {volume} {6}},\ \bibinfo {pages}
  {1} (\bibinfo {year} {2015})}\BibitemShut {NoStop}%
\bibitem [{\citenamefont {Julku}\ \emph {et~al.}(2016)\citenamefont {Julku},
  \citenamefont {Peotta}, \citenamefont {Vanhala}, \citenamefont {Kim},\ and\
  \citenamefont {T\"orm\"a}}]{PhysRevLett.117.045303}%
  \BibitemOpen
  \bibfield  {author} {\bibinfo {author} {\bibfnamefont {A.}~\bibnamefont
  {Julku}}, \bibinfo {author} {\bibfnamefont {S.}~\bibnamefont {Peotta}},
  \bibinfo {author} {\bibfnamefont {T.~I.}\ \bibnamefont {Vanhala}}, \bibinfo
  {author} {\bibfnamefont {D.-H.}\ \bibnamefont {Kim}},\ and\ \bibinfo {author}
  {\bibfnamefont {P.}~\bibnamefont {T\"orm\"a}},\ }\href
  {https://doi.org/10.1103/PhysRevLett.117.045303} {\bibfield  {journal}
  {\bibinfo  {journal} {Phys. Rev. Lett.}\ }\textbf {\bibinfo {volume} {117}},\
  \bibinfo {pages} {045303} (\bibinfo {year} {2016})}\BibitemShut {NoStop}%
\bibitem [{\citenamefont {Bleu}\ \emph {et~al.}(2018)\citenamefont {Bleu},
  \citenamefont {Malpuech}, \citenamefont {Gao},\ and\ \citenamefont
  {Solnyshkov}}]{PhysRevLett.121.020401}%
  \BibitemOpen
  \bibfield  {author} {\bibinfo {author} {\bibfnamefont {O.}~\bibnamefont
  {Bleu}}, \bibinfo {author} {\bibfnamefont {G.}~\bibnamefont {Malpuech}},
  \bibinfo {author} {\bibfnamefont {Y.}~\bibnamefont {Gao}},\ and\ \bibinfo
  {author} {\bibfnamefont {D.~D.}\ \bibnamefont {Solnyshkov}},\ }\href
  {https://doi.org/10.1103/PhysRevLett.121.020401} {\bibfield  {journal}
  {\bibinfo  {journal} {Phys. Rev. Lett.}\ }\textbf {\bibinfo {volume} {121}},\
  \bibinfo {pages} {020401} (\bibinfo {year} {2018})}\BibitemShut {NoStop}%
\bibitem [{\citenamefont {Gao}\ and\ \citenamefont
  {Xiao}(2019)}]{PhysRevLett.122.227402}%
  \BibitemOpen
  \bibfield  {author} {\bibinfo {author} {\bibfnamefont {Y.}~\bibnamefont
  {Gao}}\ and\ \bibinfo {author} {\bibfnamefont {D.}~\bibnamefont {Xiao}},\
  }\href {https://doi.org/10.1103/PhysRevLett.122.227402} {\bibfield  {journal}
  {\bibinfo  {journal} {Phys. Rev. Lett.}\ }\textbf {\bibinfo {volume} {122}},\
  \bibinfo {pages} {227402} (\bibinfo {year} {2019})}\BibitemShut {NoStop}%
\bibitem [{\citenamefont {T{\"o}rm{\"a}}\ \emph {et~al.}(2021)\citenamefont
  {T{\"o}rm{\"a}}, \citenamefont {Peotta},\ and\ \citenamefont
  {Bernevig}}]{torma2021superfluidity}%
  \BibitemOpen
  \bibfield  {author} {\bibinfo {author} {\bibfnamefont {P.}~\bibnamefont
  {T{\"o}rm{\"a}}}, \bibinfo {author} {\bibfnamefont {S.}~\bibnamefont
  {Peotta}},\ and\ \bibinfo {author} {\bibfnamefont {B.~A.}\ \bibnamefont
  {Bernevig}},\ }\href@noop {} {\bibfield  {journal} {\bibinfo  {journal}
  {arXiv preprint arXiv:2111.00807}\ } (\bibinfo {year} {2021})}\BibitemShut
  {NoStop}%
\bibitem [{\citenamefont {Gao}\ \emph {et~al.}(2014)\citenamefont {Gao},
  \citenamefont {Yang},\ and\ \citenamefont {Niu}}]{PhysRevLett.112.166601}%
  \BibitemOpen
  \bibfield  {author} {\bibinfo {author} {\bibfnamefont {Y.}~\bibnamefont
  {Gao}}, \bibinfo {author} {\bibfnamefont {S.~A.}\ \bibnamefont {Yang}},\ and\
  \bibinfo {author} {\bibfnamefont {Q.}~\bibnamefont {Niu}},\ }\href
  {https://doi.org/10.1103/PhysRevLett.112.166601} {\bibfield  {journal}
  {\bibinfo  {journal} {Phys. Rev. Lett.}\ }\textbf {\bibinfo {volume} {112}},\
  \bibinfo {pages} {166601} (\bibinfo {year} {2014})}\BibitemShut {NoStop}%
\bibitem [{\citenamefont {Gao}\ \emph {et~al.}(2015)\citenamefont {Gao},
  \citenamefont {Yang},\ and\ \citenamefont {Niu}}]{PhysRevB.91.214405}%
  \BibitemOpen
  \bibfield  {author} {\bibinfo {author} {\bibfnamefont {Y.}~\bibnamefont
  {Gao}}, \bibinfo {author} {\bibfnamefont {S.~A.}\ \bibnamefont {Yang}},\ and\
  \bibinfo {author} {\bibfnamefont {Q.}~\bibnamefont {Niu}},\ }\href
  {https://doi.org/10.1103/PhysRevB.91.214405} {\bibfield  {journal} {\bibinfo
  {journal} {Phys. Rev. B}\ }\textbf {\bibinfo {volume} {91}},\ \bibinfo
  {pages} {214405} (\bibinfo {year} {2015})}\BibitemShut {NoStop}%
\bibitem [{\citenamefont {Gao}(2019)}]{gao2019semiclassical}%
  \BibitemOpen
  \bibfield  {author} {\bibinfo {author} {\bibfnamefont {Y.}~\bibnamefont
  {Gao}},\ }\href {https://doi.org/10.1007/s11467-019-0887-2} {\bibfield
  {journal} {\bibinfo  {journal} {Front. Phys.}\ }\textbf {\bibinfo {volume}
  {14}},\ \bibinfo {pages} {1} (\bibinfo {year} {2019})}\BibitemShut {NoStop}%
\bibitem [{\citenamefont {Liu}\ \emph {et~al.}(2021)\citenamefont {Liu},
  \citenamefont {Zhao}, \citenamefont {Huang}, \citenamefont {Wu},
  \citenamefont {Sheng}, \citenamefont {Xiao},\ and\ \citenamefont
  {Yang}}]{PhysRevLett.127.277202}%
  \BibitemOpen
  \bibfield  {author} {\bibinfo {author} {\bibfnamefont {H.}~\bibnamefont
  {Liu}}, \bibinfo {author} {\bibfnamefont {J.}~\bibnamefont {Zhao}}, \bibinfo
  {author} {\bibfnamefont {Y.-X.}\ \bibnamefont {Huang}}, \bibinfo {author}
  {\bibfnamefont {W.}~\bibnamefont {Wu}}, \bibinfo {author} {\bibfnamefont
  {X.-L.}\ \bibnamefont {Sheng}}, \bibinfo {author} {\bibfnamefont
  {C.}~\bibnamefont {Xiao}},\ and\ \bibinfo {author} {\bibfnamefont {S.~A.}\
  \bibnamefont {Yang}},\ }\href
  {https://doi.org/10.1103/PhysRevLett.127.277202} {\bibfield  {journal}
  {\bibinfo  {journal} {Phys. Rev. Lett.}\ }\textbf {\bibinfo {volume} {127}},\
  \bibinfo {pages} {277202} (\bibinfo {year} {2021})}\BibitemShut {NoStop}%
\bibitem [{\citenamefont {Rossi}(2021)}]{rossi2021quantum}%
  \BibitemOpen
  \bibfield  {author} {\bibinfo {author} {\bibfnamefont {E.}~\bibnamefont
  {Rossi}},\ }\href@noop {} {\bibfield  {journal} {\bibinfo  {journal} {Current
  Opinion in Solid State and Materials Science}\ }\textbf {\bibinfo {volume}
  {25}},\ \bibinfo {pages} {100952} (\bibinfo {year} {2021})}\BibitemShut
  {NoStop}%
\bibitem [{\citenamefont {Holder}\ \emph {et~al.}(2020)\citenamefont {Holder},
  \citenamefont {Kaplan},\ and\ \citenamefont
  {Yan}}]{PhysRevResearch.2.033100}%
  \BibitemOpen
  \bibfield  {author} {\bibinfo {author} {\bibfnamefont {T.}~\bibnamefont
  {Holder}}, \bibinfo {author} {\bibfnamefont {D.}~\bibnamefont {Kaplan}},\
  and\ \bibinfo {author} {\bibfnamefont {B.}~\bibnamefont {Yan}},\ }\href
  {https://doi.org/10.1103/PhysRevResearch.2.033100} {\bibfield  {journal}
  {\bibinfo  {journal} {Phys. Rev. Research}\ }\textbf {\bibinfo {volume}
  {2}},\ \bibinfo {pages} {033100} (\bibinfo {year} {2020})}\BibitemShut
  {NoStop}%
\bibitem [{\citenamefont {Xiao}\ \emph {et~al.}(2010)\citenamefont {Xiao},
  \citenamefont {Chang},\ and\ \citenamefont {Niu}}]{RevModPhys.82.1959}%
  \BibitemOpen
  \bibfield  {author} {\bibinfo {author} {\bibfnamefont {D.}~\bibnamefont
  {Xiao}}, \bibinfo {author} {\bibfnamefont {M.-C.}\ \bibnamefont {Chang}},\
  and\ \bibinfo {author} {\bibfnamefont {Q.}~\bibnamefont {Niu}},\ }\href
  {https://doi.org/10.1103/RevModPhys.82.1959} {\bibfield  {journal} {\bibinfo
  {journal} {Rev. Mod. Phys.}\ }\textbf {\bibinfo {volume} {82}},\ \bibinfo
  {pages} {1959} (\bibinfo {year} {2010})}\BibitemShut {NoStop}%
\bibitem [{\citenamefont {Duval}\ \emph {et~al.}(2006)\citenamefont {Duval},
  \citenamefont {Horv{\'a}th}, \citenamefont {Horvathy}, \citenamefont
  {Martina},\ and\ \citenamefont {Stichel}}]{duval2006berry}%
  \BibitemOpen
  \bibfield  {author} {\bibinfo {author} {\bibfnamefont {C.}~\bibnamefont
  {Duval}}, \bibinfo {author} {\bibfnamefont {Z.}~\bibnamefont {Horv{\'a}th}},
  \bibinfo {author} {\bibfnamefont {P.~A.}\ \bibnamefont {Horvathy}}, \bibinfo
  {author} {\bibfnamefont {L.}~\bibnamefont {Martina}},\ and\ \bibinfo {author}
  {\bibfnamefont {P.}~\bibnamefont {Stichel}},\ }\href@noop {} {\bibfield
  {journal} {\bibinfo  {journal} {Modern Physics Letters B}\ }\textbf {\bibinfo
  {volume} {20}},\ \bibinfo {pages} {373} (\bibinfo {year} {2006})}\BibitemShut
  {NoStop}%
\bibitem [{\citenamefont {Xiao}\ \emph {et~al.}(2005)\citenamefont {Xiao},
  \citenamefont {Shi},\ and\ \citenamefont {Niu}}]{PhysRevLett.95.137204}%
  \BibitemOpen
  \bibfield  {author} {\bibinfo {author} {\bibfnamefont {D.}~\bibnamefont
  {Xiao}}, \bibinfo {author} {\bibfnamefont {J.}~\bibnamefont {Shi}},\ and\
  \bibinfo {author} {\bibfnamefont {Q.}~\bibnamefont {Niu}},\ }\href
  {https://doi.org/10.1103/PhysRevLett.95.137204} {\bibfield  {journal}
  {\bibinfo  {journal} {Phys. Rev. Lett.}\ }\textbf {\bibinfo {volume} {95}},\
  \bibinfo {pages} {137204} (\bibinfo {year} {2005})}\BibitemShut {NoStop}%
\bibitem [{\citenamefont {Ma}\ \emph {et~al.}(2019)\citenamefont {Ma},
  \citenamefont {Jiang}, \citenamefont {Liu},\ and\ \citenamefont
  {Xie}}]{PhysRevB.99.115121}%
  \BibitemOpen
  \bibfield  {author} {\bibinfo {author} {\bibfnamefont {D.}~\bibnamefont
  {Ma}}, \bibinfo {author} {\bibfnamefont {H.}~\bibnamefont {Jiang}}, \bibinfo
  {author} {\bibfnamefont {H.}~\bibnamefont {Liu}},\ and\ \bibinfo {author}
  {\bibfnamefont {X.~C.}\ \bibnamefont {Xie}},\ }\href
  {https://doi.org/10.1103/PhysRevB.99.115121} {\bibfield  {journal} {\bibinfo
  {journal} {Phys. Rev. B}\ }\textbf {\bibinfo {volume} {99}},\ \bibinfo
  {pages} {115121} (\bibinfo {year} {2019})}\BibitemShut {NoStop}%
\bibitem [{\citenamefont {Page}(1987)}]{PhysRevA.36.3479}%
  \BibitemOpen
  \bibfield  {author} {\bibinfo {author} {\bibfnamefont {D.~N.}\ \bibnamefont
  {Page}},\ }\href {https://doi.org/10.1103/PhysRevA.36.3479} {\bibfield
  {journal} {\bibinfo  {journal} {Phys. Rev. A}\ }\textbf {\bibinfo {volume}
  {36}},\ \bibinfo {pages} {3479} (\bibinfo {year} {1987})}\BibitemShut
  {NoStop}%
\bibitem [{\citenamefont {Bengtsson}\ and\ \citenamefont
  {{\.Z}yczkowski}(2017)}]{bengtsson2017geometry}%
  \BibitemOpen
  \bibfield  {author} {\bibinfo {author} {\bibfnamefont {I.}~\bibnamefont
  {Bengtsson}}\ and\ \bibinfo {author} {\bibfnamefont {K.}~\bibnamefont
  {{\.Z}yczkowski}},\ }\href@noop {} {\emph {\bibinfo {title} {Geometry of
  quantum states: an introduction to quantum entanglement}}}\ (\bibinfo
  {publisher} {Cambridge university press},\ \bibinfo {year}
  {2017})\BibitemShut {NoStop}%
\bibitem [{\citenamefont {Cheng}(2010)}]{cheng2010quantum}%
  \BibitemOpen
  \bibfield  {author} {\bibinfo {author} {\bibfnamefont {R.}~\bibnamefont
  {Cheng}},\ }\href@noop {} {\bibfield  {journal} {\bibinfo  {journal} {arXiv
  preprint arXiv:1012.1337}\ } (\bibinfo {year} {2010})}\BibitemShut {NoStop}%
\bibitem [{sup()}]{supp}%
  \BibitemOpen
  \href@noop {} {}\bibinfo {howpublished}
  {\url{URL_will_be_inserted_by_publisher}},\ \bibinfo {note} {see Supplemental
  Material at ... for (i) detailed derivations of the intrinsic nonlinear NHT
  current, (ii) symmetry constrains on $\sigma_{\text{NHT}}$.}\BibitemShut
  {Stop}%
\bibitem [{\citenamefont {Ge}\ \emph {et~al.}(2020)\citenamefont {Ge},
  \citenamefont {Ma}, \citenamefont {Liu}, \citenamefont {Wang}, \citenamefont
  {Li}, \citenamefont {Luo}, \citenamefont {Luo}, \citenamefont {Xing},
  \citenamefont {Yan}, \citenamefont {Mandrus} \emph
  {et~al.}}]{ge2020unconventional}%
  \BibitemOpen
  \bibfield  {author} {\bibinfo {author} {\bibfnamefont {J.}~\bibnamefont
  {Ge}}, \bibinfo {author} {\bibfnamefont {D.}~\bibnamefont {Ma}}, \bibinfo
  {author} {\bibfnamefont {Y.}~\bibnamefont {Liu}}, \bibinfo {author}
  {\bibfnamefont {H.}~\bibnamefont {Wang}}, \bibinfo {author} {\bibfnamefont
  {Y.}~\bibnamefont {Li}}, \bibinfo {author} {\bibfnamefont {J.}~\bibnamefont
  {Luo}}, \bibinfo {author} {\bibfnamefont {T.}~\bibnamefont {Luo}}, \bibinfo
  {author} {\bibfnamefont {Y.}~\bibnamefont {Xing}}, \bibinfo {author}
  {\bibfnamefont {J.}~\bibnamefont {Yan}}, \bibinfo {author} {\bibfnamefont
  {D.}~\bibnamefont {Mandrus}}, \emph {et~al.},\ }\href@noop {} {\bibfield
  {journal} {\bibinfo  {journal} {National Science Review}\ }\textbf {\bibinfo
  {volume} {7}},\ \bibinfo {pages} {1879} (\bibinfo {year} {2020})}\BibitemShut
  {NoStop}%
\bibitem [{\citenamefont {Arnold}\ \emph {et~al.}(2016)\citenamefont {Arnold},
  \citenamefont {Naumann}, \citenamefont {Wu}, \citenamefont {Sun},
  \citenamefont {Schmidt}, \citenamefont {Borrmann}, \citenamefont {Felser},
  \citenamefont {Yan},\ and\ \citenamefont
  {Hassinger}}]{PhysRevLett.117.146401}%
  \BibitemOpen
  \bibfield  {author} {\bibinfo {author} {\bibfnamefont {F.}~\bibnamefont
  {Arnold}}, \bibinfo {author} {\bibfnamefont {M.}~\bibnamefont {Naumann}},
  \bibinfo {author} {\bibfnamefont {S.-C.}\ \bibnamefont {Wu}}, \bibinfo
  {author} {\bibfnamefont {Y.}~\bibnamefont {Sun}}, \bibinfo {author}
  {\bibfnamefont {M.}~\bibnamefont {Schmidt}}, \bibinfo {author} {\bibfnamefont
  {H.}~\bibnamefont {Borrmann}}, \bibinfo {author} {\bibfnamefont
  {C.}~\bibnamefont {Felser}}, \bibinfo {author} {\bibfnamefont
  {B.}~\bibnamefont {Yan}},\ and\ \bibinfo {author} {\bibfnamefont
  {E.}~\bibnamefont {Hassinger}},\ }\href
  {https://doi.org/10.1103/PhysRevLett.117.146401} {\bibfield  {journal}
  {\bibinfo  {journal} {Phys. Rev. Lett.}\ }\textbf {\bibinfo {volume} {117}},\
  \bibinfo {pages} {146401} (\bibinfo {year} {2016})}\BibitemShut {NoStop}%
\bibitem [{\citenamefont {Li}\ \emph {et~al.}(2021)\citenamefont {Li},
  \citenamefont {Heinonen}, \citenamefont {Burkov},\ and\ \citenamefont
  {Zhang}}]{PhysRevB.103.045105}%
  \BibitemOpen
  \bibfield  {author} {\bibinfo {author} {\bibfnamefont {R.-H.}\ \bibnamefont
  {Li}}, \bibinfo {author} {\bibfnamefont {O.~G.}\ \bibnamefont {Heinonen}},
  \bibinfo {author} {\bibfnamefont {A.~A.}\ \bibnamefont {Burkov}},\ and\
  \bibinfo {author} {\bibfnamefont {S.~S.-L.}\ \bibnamefont {Zhang}},\ }\href
  {https://doi.org/10.1103/PhysRevB.103.045105} {\bibfield  {journal} {\bibinfo
   {journal} {Phys. Rev. B}\ }\textbf {\bibinfo {volume} {103}},\ \bibinfo
  {pages} {045105} (\bibinfo {year} {2021})}\BibitemShut {NoStop}%
\bibitem [{\citenamefont {Zhang}\ \emph {et~al.}(2018)\citenamefont {Zhang},
  \citenamefont {Sun},\ and\ \citenamefont {Yan}}]{PhysRevB.97.041101}%
  \BibitemOpen
  \bibfield  {author} {\bibinfo {author} {\bibfnamefont {Y.}~\bibnamefont
  {Zhang}}, \bibinfo {author} {\bibfnamefont {Y.}~\bibnamefont {Sun}},\ and\
  \bibinfo {author} {\bibfnamefont {B.}~\bibnamefont {Yan}},\ }\href
  {https://doi.org/10.1103/PhysRevB.97.041101} {\bibfield  {journal} {\bibinfo
  {journal} {Phys. Rev. B}\ }\textbf {\bibinfo {volume} {97}},\ \bibinfo
  {pages} {041101} (\bibinfo {year} {2018})}\BibitemShut {NoStop}%
\bibitem [{\citenamefont {Zeng}\ \emph {et~al.}(2021)\citenamefont {Zeng},
  \citenamefont {Nandy},\ and\ \citenamefont {Tewari}}]{PhysRevB.103.245119}%
  \BibitemOpen
  \bibfield  {author} {\bibinfo {author} {\bibfnamefont {C.}~\bibnamefont
  {Zeng}}, \bibinfo {author} {\bibfnamefont {S.}~\bibnamefont {Nandy}},\ and\
  \bibinfo {author} {\bibfnamefont {S.}~\bibnamefont {Tewari}},\ }\href
  {https://doi.org/10.1103/PhysRevB.103.245119} {\bibfield  {journal} {\bibinfo
   {journal} {Phys. Rev. B}\ }\textbf {\bibinfo {volume} {103}},\ \bibinfo
  {pages} {245119} (\bibinfo {year} {2021})}\BibitemShut {NoStop}%
\bibitem [{\citenamefont {Matsyshyn}\ and\ \citenamefont
  {Sodemann}(2019)}]{PhysRevLett.123.246602}%
  \BibitemOpen
  \bibfield  {author} {\bibinfo {author} {\bibfnamefont {O.}~\bibnamefont
  {Matsyshyn}}\ and\ \bibinfo {author} {\bibfnamefont {I.}~\bibnamefont
  {Sodemann}},\ }\href {https://doi.org/10.1103/PhysRevLett.123.246602}
  {\bibfield  {journal} {\bibinfo  {journal} {Phys. Rev. Lett.}\ }\textbf
  {\bibinfo {volume} {123}},\ \bibinfo {pages} {246602} (\bibinfo {year}
  {2019})}\BibitemShut {NoStop}%
\bibitem [{\citenamefont {Nandy}\ \emph {et~al.}(2021)\citenamefont {Nandy},
  \citenamefont {Zeng},\ and\ \citenamefont {Tewari}}]{PhysRevB.104.205124}%
  \BibitemOpen
  \bibfield  {author} {\bibinfo {author} {\bibfnamefont {S.}~\bibnamefont
  {Nandy}}, \bibinfo {author} {\bibfnamefont {C.}~\bibnamefont {Zeng}},\ and\
  \bibinfo {author} {\bibfnamefont {S.}~\bibnamefont {Tewari}},\ }\href
  {https://doi.org/10.1103/PhysRevB.104.205124} {\bibfield  {journal} {\bibinfo
   {journal} {Phys. Rev. B}\ }\textbf {\bibinfo {volume} {104}},\ \bibinfo
  {pages} {205124} (\bibinfo {year} {2021})}\BibitemShut {NoStop}%
\bibitem [{\citenamefont {Huang}\ \emph {et~al.}(2022)\citenamefont {Huang},
  \citenamefont {Feng}, \citenamefont {Wang}, \citenamefont {Xiao},\ and\
  \citenamefont {Yang}}]{huang2022intrinsic}%
  \BibitemOpen
  \bibfield  {author} {\bibinfo {author} {\bibfnamefont {Y.-X.}\ \bibnamefont
  {Huang}}, \bibinfo {author} {\bibfnamefont {X.}~\bibnamefont {Feng}},
  \bibinfo {author} {\bibfnamefont {H.}~\bibnamefont {Wang}}, \bibinfo {author}
  {\bibfnamefont {C.}~\bibnamefont {Xiao}},\ and\ \bibinfo {author}
  {\bibfnamefont {S.~A.}\ \bibnamefont {Yang}},\ }\href@noop {} {\bibfield
  {journal} {\bibinfo  {journal} {arXiv preprint arXiv:2208.03639}\ } (\bibinfo
  {year} {2022})}\BibitemShut {NoStop}%
\end{thebibliography}
%

\end{document}